\documentclass[pra,twocolumn,superscriptaddress,preprintnumbers,amsmath,amssymb,nofootinbib,floatfix,longbibliography]{revtex4-2}
\usepackage{graphicx,bm}
\usepackage{float}
\usepackage{subfigure}
\usepackage{amsmath}
\usepackage{amsfonts}
\usepackage{tikz}
\usepackage{relsize}
\usetikzlibrary{shapes.geometric, calc}
\usepackage{inputenc}
\usepackage{amsthm, amssymb,bbm,bm}

\usepackage[normalem]{ulem}

\begin{document}
\title{Integrated quantum polariton interferometry}
\author{Davide Nigro}
\affiliation{Dipartimento di Fisica, Universit\`{a} di Pavia, via Bassi 6, I-27100 Pavia, Italy}
\author{Vincenzo D'Ambrosio}
\affiliation{Dipartimento di Fisica, Universit\`{a} di Napoli Federico II, Complesso Universitario di Monte Sant'Angelo, Via Cintia, 80126 Napoli, Italy}
\author{Daniele Sanvitto}
\affiliation{CNR NANOTEC, Institute of Nanotechnology, Campus Ecotekne, Via Monteroni, 73100, Lecce, Italy}
\author{Dario Gerace}
\affiliation{Dipartimento di Fisica, Universit\`{a} di Pavia, via Bassi 6, I-27100 Pavia, Italy}

\begin{abstract}
{Exciton-polaritons are hybrid elementary excitations of light and matter that, thanks to their nonlinear properties, enable a plethora of physical phenomena ranging from room temperature condensation to superfluidity. While polaritons are usually exploited in high density regime, evidence of quantum correlations at the level of few excitations has been recently reported, thus suggesting the possibility of using these systems for quantum information purposes. Here we show that integrated circuits of propagating single polaritons can be arranged to build deterministic quantum logic gates in which the two-particle interaction energy plays a crucial role. Besides showing their prospective potential for photonic quantum computation, we also show that these systems can be exploited for metrology purposes, as for instance to precisely measure the magnitude of the polariton-polariton interaction at the two-body level. In general, our results introduce a novel paradigm for the development of practical quantum polaritonic devices, in which the effective interaction between single polaritonic qubits may provide a unique tool for future quantum technologies.}
\end{abstract}
\maketitle

\section{Introduction}

Exciton-polaritons have been emerging as an incredibly promising platform to explore fundamental physics effects in an analog solid state scenario, such as Bose-Einstein condensation \cite{Kasprzak2006,Balili2007}, superfluidity \cite{Amo2009}, soliton propagation\cite{Amo2011}, spontaneous formation of vortices \cite{Lagoudakis2008}, Josephson oscillations and self-trapping \cite{Abbarchi2013}, analog gravity \cite{HaiSon2015prl}. In parallel, their remarkable properties have been exploited to deliver more application-oriented devices, such as all-optical transistors \cite{Ballarini2013}, resonant tunnel diodes \cite{HaiSon2013prl}, Mach-Zehnder interferometers \cite{Sturm2014}, routers \cite{Marsault2015}, couplers \cite{Beierlein2021}, or ultra-low-threshold lasers \cite{Azzini2011}, just to mention a few. Polariton integrated circuits have been proposed in view of mimicking the key functionalities of electronic circuits \cite{Liew2010} or to implement artificial neural networks \cite{Liew2008}. Polariton excitations are characterized by a coherent superposition of two very different fields, the electromagnetic (photon) field and an optically active polarization (the exciton), i.e., a collective crystal excitation made of a bound electron-hole pair \cite{Andreani1990}. As such, the polariton field inherits the small effective mass (in the order of $10^{-5}$ relative to the free electron mass) of bound photonic modes, e.g., in a planar microcavity or waveguide \cite{Kavokin2008,Sanvitto2016}. In addition, it also inherits the large nonlinearity (comparatively, estimated to be about 4 orders of magnitude larger than bulk silicon, for example) derived from the Coulomb interaction between excitons \cite{Ciuti1998,Tassone1999}. At high excitation density (but well below the saturation regime), this allows to formally describe the polariton mean field in terms of a non-equilibrium Gross-Pitaevskii equation of motion, which has been successfully applied to describe most of the quantum fluid phenomenology  observed, so far \cite{Carusotto-Ciuti2013}. This is a further confirmation that confined polaritons can be effectively treated as an out-of-equilibrium gas of weakly interacting bosons, whose steady state inevitably depends on the balance between driving and losses in the system. Recently, exciton-polariton condensates have also been proposed for quantum information processing, with the proposal of encoding qubits into the quantum fluctuations on top of the condensate \cite{Ghosh2020}, or in the polariton spin degrees of freedom \cite{Solnyshkov2015}.

\begin{figure}[b]
    \centering
   \includegraphics[scale=0.5]{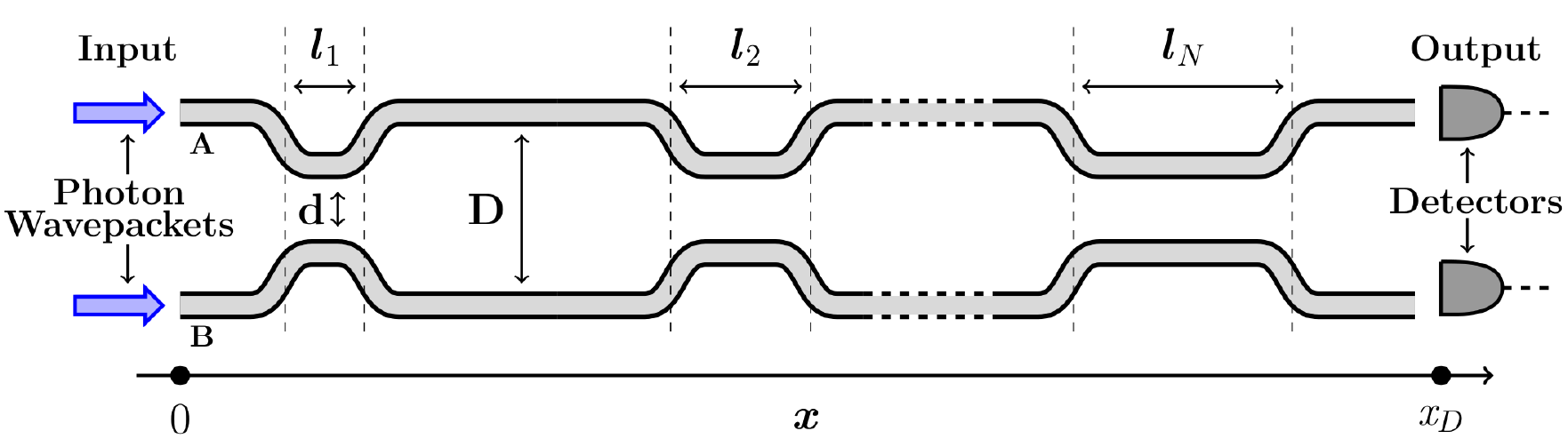}
    \caption{General scheme of an integrated polariton interferometer. It is assumed that polaritons are created at $x=0$ and many-body interactions are then detected at $x_D$ by measuring the degree of correlation of photons emerging from the two channels. In this work, we will assume that each input is a single photon state. The device consists of a sequence of $N$ interaction regions of length $l_j$ ($j=1,\,2,\,\cdots N$) where the two waveguides (labeled as $A$ and $B$) run in parallel at a distance equal to $d$ so that they are evanescently coupled. Two consecutive interaction regions are separated by a free-propagation region, that is a portion of the device where the distance between the two waveguides is $D\gg d$ and the coupling between the two waveguides is negligible.}
    \label{fig:generic_device}
\end{figure}

More recent experiments have been exploring the possibility to excite the quantized polariton field in confined geometries at the level of single or few excitation quanta \cite{Cuevas2018,suarez-forero_hydrodynamics_2020}. In this regime, the mean field treatment does not provide an accurate theoretical description, calling for a more appropriate open quantum system approach to describe strongly interacting polaritons \cite{Carusotto-Ciuti2013}. Recently, evidence  of quantum correlations at the level of few polariton excitations has been reported \cite{Matutano-emergence-2019,Delteil-blockade-2019}, which has triggered our motivation to provide an in-depth theoretical analysis of integrated polariton circuits in the quantum regime, with particular attention to the differences with respect to photonic integrated circuits {and to the unique possibilities enabled by} nonlinearities introduced by polariton interactions. In our scenario, such \textit{quantum polariton integrated circuits} (QPIC) can be assumed as complex networks resulting from the combination of coupled waveguiding channels and interferometers, in which polaritons can be injected and propagate while interacting at the level of few quanta. 
{W}e will hereby focus on a realistic experimental configuration in which single photons (generated, e.g., from a pair of quantum emitters) are coupled with a given efficiency onto a semiconductor device in which polariton excitations exist and can be guided on chip \cite{LopezCarreno2015}. In such a situation, single photons are converted into single polariton wave packets that retain the quantum radiation field coherence properties, as recently shown \cite{Cuevas2018,suarez-forero_hydrodynamics_2020}, and can then be interfered through evanescently coupled waveguides alternated with regions of free propagation, as illustrated in Fig.~\ref{fig:generic_device}. 

We notice that the precise characterization of a linear device having an arbitrary large number $n$ of propagation channels with $m<n$ single-photon inputs is already a challenging task to be solved, even by means of advanced numerical tools. Therefore, this problem has  recently attracted a great deal of attention, both in theory and experiments, aiming at exploiting quantum resources to demonstrate the solution of a problem that no classical computer can solve in a feasible amount of time, the so called quantum supremacy \cite{BosonSamplingReviewBrod,BosonSamplingAaronson,BosonSamplingEXPPrl,Zhong2020supremacy}. Interestingly, it has been conjectured in this context that the presence of nonlinearities may be beneficial to reach quantum supremacy,  possibly in an integrated configuration \cite{Sciarrino2021SciBull}. 
More in general, nonlinearities can be a crucial resource in the quest for the development of photonic quantum gates for quantum information processing. Indeed, while original proposals only employing linear optical elements lead to a non-deterministic computing scenario \cite{Knill_Nature2001_LOQC,Ralph_PRA_2002}, novel ideas aimed at developing deterministic quantum gates for photonic quantum computing, based on effective photonic interactions, have been put forward \cite{Cala19,Englund_PRL_2020,Li_PRApp_2020}. The realization of the latter paradigm would represent a significant milestone for quantum technologies, already in near-term devices \cite{Preskill2018NISQ}. 
{Within this generalized context, we show that it is possible to realize nonlinear quantum devices by exploiting  polariton-polariton interactions that are naturally present in QPIC. In particular, we introduce a scheme to implement a fundamental two-qubits quantum gate, specifically the  $\sqrt{\mbox{SWAP}'}$ gate introduced in Ref.~\cite{Franson2004}. Such a gate can be used in combination with single qubit operations to implement the CNOT gate (thus, \emph{any} quantum gate \cite{Nielsen_Chuang}). Therefore our results demonstrate that QPICs represent a promising platform for the development of integrated photonic devices for universal quantum computing, going beyond currently accepted paradigms based on linear optical elements. The novel functionalities introduced by QPICs  can then be exploited in addition to (or in combination with) the usual linear optical interferometers employed in conventional photonic circuits (PCs) \cite{Politi2008,Crespi2013,Metcalf2013}.  }

{Noticeably, the QPICs here introduced are shown to hold promise also in quantum metrology. In particular, they potentially allow to measure with high precision the two-polariton nonlinear shift, a quantity that has only been indirectly estimated so far. In fact, it is worth pointing out that there has been a longstanding debate to quantitatively measure the polariton-polariton nonlinearity \cite{Ferrier2011}, and only recently the corresponding order of magnitude has been more precisely inferred from photon-photon correlation measurements for three-dimensional polariton confinement \cite{Delteil-blockade-2019}. Moreover, an increase in the magnitude of this parameter, induced by a strong electric field has been recently reported \cite{Rosenberg2018,Suarez-Forero2021}. However, despite the considerable interest in knowing the precise value of the polariton interaction energy at the two-body level, no clear and quantitatively accurate answer to this question exists, yet. Essentially, this is because most of these measurements are performed in the regime of large polariton density, typically treated at the mean field level, and whose nonlinear behavior is fully characterized by the product of the inter-particle interaction energy times the particle density. In practice, disentangling these two contributions is at the origin of large uncertainties in the determination of the inter-particle contribution alone. Here we propose a new way of giving a precise characterization of the polariton-polariton interaction based on polariton Fock states interference in QPIC architecture.}

{More generally speaking, the framework} here developed may pave the way for the {realization} of novel quantum technologies enabled by polariton-polariton interactions, with applications ranging from quantum metrology, simulations, and sensing, to quantum information processing and computing. 


\section{Theoretical Framework }\label{sec:theory}
{We start by briefly outlining the theoretical model that will be exploited to describe polariton-polariton interactions at the two-particle level in the QPIC} of Fig.~\ref{fig:generic_device}. In what follows, we will only consider the Hamiltonian unitary evolution, which {already} allows to grasp the effects of two-particle interactions on quantum correlations. In fact, {In fact long lived polaritonic excitations \cite{Nelsen2013dissipationless}, and propagation lengths in the order of 400 $\mu$m  \cite{suarez-forero_hydrodynamics_2020},  justify our unitary approach if restricted to devices with limited length.}\\
However, a quantitative description of the effects of particle losses, unavoidably present in realistic experimental situations, 
is given in Supplementary information (SI),
{showing how such incoherent processes do not affect the output photon statistics derived in the Hamiltonian evolution. } 

\subsection{Modeling the time evolution}
The effective Hamiltonian model used to describe the propagation of polaritons in the two co-directional waveguides sketched in Fig. \ref{fig:generic_device} reads ($\hbar=1$)
\begin{equation}\label{eq:model_Hamiltonian}
\mathcal{\hat{H}}=\sum_{k}\left[\mathcal{\hat{H}}^{a}_{k} + \mathcal{\hat{H}}^{b}_{k}+J(x) \left(\hat{a}^{\dagger}_k \hat{b}_k+\hat{b}^{\dagger}_k \hat{a}_k\right)\right] \, ,
\end{equation}
in which $\hat{a}^{(\dagger)}_k$ and $\hat{b}^{(\dagger)}_k$ denote the annihilation (creation) operators of a polariton with wavevector $k$ in the waveguide $A$ and $B$ respectively, and     $\mathcal{\hat{H}}^{\sigma}_{k}=\omega_k \hat{\sigma}^{\dagger}_k \hat{\sigma}_k + U_k\,\hat{\sigma}^{\dagger}_k \hat{\sigma}^{\dagger}_k \hat{\sigma}_k \hat{\sigma}_k$ ($\sigma=a,\,b$) represents the Hamiltonian of each waveguide in the lower polariton branch. A detailed derivation of this model from the coupled exciton-photon fields is reported in the SI. 
In particular, the interaction energy $U_k$ accounts for the two-body interaction arising from the exciton fraction in polaritons at given wave vector $k$. The last term in Eq. \ref{eq:model_Hamiltonian} describes hopping between the two waveguides, which is due to the evanescent coupling between the photonic fractions of the co-propagating polaritons in $A$ and $B$ channels, respectively. This is formally equivalent to a beam splitter Hamiltonian in quantum optics. We can generically assume that the tunnel coupling energy, $J(x)$, is a space-dependent parameter only determined by the physical distance between the two channels at position $x$ along the propagation direction. Furthermore, since the evanescent fields in each waveguide decay exponentially, hopping is expected to be suppressed when increasing the separation between the two channels, while being effectively present only in the regions where the two waveguides are close enough. As a consequence, it is reasonable to assume $J(x)$ to be non-zero and equal to a constant value, $J>0$, only when the distance is $d$, and zero elsewhere.\\
In our scenario, we envision single polariton quanta to be created in the integrated device by shining single photons on the leftmost side of the device, at $x=0$, and their properties (for instance, polariton-polariton correlations) are subsequently measured by detecting photons emerging on the rightmost side of the apparatus, at $x=x_D$. In particular, we consider the initial state as a product state, that is
\begin{equation}
    \vert \psi (t=0)\rangle = \vert \psi_{A}(t=0)\rangle \otimes \vert \psi_{B}(t=0)\rangle, 
\end{equation}
with $\vert \psi_{A}(t=0)\rangle$ and $\vert \psi_{B}(t=0)\rangle$ describing the initial state of the waveguide $A$ and $B$ respectively. Although at $t=0$ the quantum system is in a product state, at $t>0$ it will be in a superposition of different quantum states. In addition to the obvious dependence on geometrical aspects, such as the length of the interaction regions ($\{l_j\}$) and the separation between the waveguides, the degree of correlation present in $\vert \psi (t) \rangle$ does depend on several factors, ranging from the number of initial photons injected into the system, to their temporal and spatial shape. In the present paper we will only consider the cases where the initial state corresponds to either (i) a single polariton, or (ii) a product of two single-polariton Fock states. Furthermore, we assume in both cases to have particles with the same wavevector, $k$. More explicitly, in what follows we consider single polariton states like the following ones
\begin{equation}\label{eq:one_A}
   \vert \psi (t=0)\rangle =\hat{a}^{\dagger}_k \vert \Omega\rangle \equiv \vert 1^A_{k},\, 0^B_{k}\rangle, 
\end{equation}
or 
\begin{equation}\label{eq:one_B}
    \vert \psi (t=0)\rangle =\hat{b}^{\dagger}_k \vert \Omega\rangle \equiv \vert 0^{A}_k,\, 1^{B}_k\rangle,
\end{equation}
and two polariton states having the following expression
\begin{equation}\label{eq:initial_state}
     \vert \psi (t=0)\rangle =\hat{a}^{\dagger}_k\hat{b}^{\dagger}_k \vert \Omega\rangle \equiv \vert 1^{A}_k,\, 1^{B}_k\rangle,
\end{equation}
with $\vert \Omega\rangle $ denoting the polariton vacuum in each propagating channel. In the next two sections we discuss the evolution of the states defined in Eqs. \ref{eq:one_A}, \ref{eq:one_B} and \ref{eq:initial_state}.\\
\subsection{Dynamics of a single polariton state}\label{sec:single_polariton_evo}
The one polariton states in Eq. \ref{eq:one_A} and \ref{eq:one_B}
are not affected by interaction terms proportional to $U_k$, and their evolution through the QPIC in Fig. \ref{fig:generic_device} is only sensitive to the regions where $J(x)\neq 0$. Let us assume now that a polariton state with wavevector $k$ corresponds to a particle propagating at a constant speed given by its group velocity, that is 
\begin{equation}
    v_g= \left.\frac{\partial\,\omega_k}{\partial k}\right\vert_k,
\end{equation}
with $\omega_k$ being the one polariton dispersion entering in the Hamiltonian model. This implies that, in order to propagate across the $m$-th interaction region of length $l_m$, the quantum particle needs roughly a time given by
\begin{equation}
t_m=\frac{l_m}{v_g}.    
\end{equation}
In addition, by observing that such states are in one-to-one correspondence with the two polarization states of a spin-1/2 particle, that is
\begin{equation}
     \vert \uparrow\, \rangle = \vert 1^{A}_k,\, 0^{B}_k\rangle,\quad \vert \downarrow\, \rangle = \vert 0^{A}_k,\,1^{B}_k\rangle,
\end{equation}
their evolution through an interaction length $l=v_g\, t$ is prescribed by the following unitary operator
 \begin{equation}\label{eq:BS_action}
 \begin{split}
     U^{(1)}_{IR}(t)&=e^{-i \mathcal{\hat{H}} t} =\\ &=e^{-i\omega_k t}\left(\cos(J t)\mathbbm{1}-i\,\sin(J t)\sigma_x\right),
\end{split}
 \end{equation}
with $\hat{\mathcal{H}}=\omega_k \mathbbm{1}+J\sigma_x$, with $\mathbbm{1}$ and $\sigma_x$ being the identity and the Pauli-X matrices respectively. The subscript IR stands for "Interaction region".\\
In particular, notice that in correspondence of the following values of the propagation time
\begin{equation}\label{eq:BS_times}
    T^{(n)}_{dip}= \frac{\pi}{4J}(1+2\,n),
\end{equation}
with $n$ being any integer, the operator in Eq. \ref{eq:BS_action} describes a 50:50 beamsplitter. In practice, this means that if one properly chooses the interaction length $l$ and the group-velocity $v_g$ in such a way that the ratio $l/v_g$ equals $ T^{(n)}_{dip}$ for some value of $n$, then an initial single polariton state, such as those in Eqs. \ref{eq:one_A} and \ref{eq:one_B}, is mapped into their equally weighted superposition. For instance, in correspondence of $T^{(0)}_{dip}\equiv T_{dip}$, one has that 
\begin{equation}
    U^{(1)}_{IR}(T_{dip})\vert 1^{A}_k,\,0^{B}_k\rangle=\frac{\vert 1^{A}_k,\, 0^{B}_k\rangle -i \,\vert 0^{A}_k,\,1^{B}_k\rangle }{\sqrt{2}}
\end{equation}
and that
\begin{equation}
    U^{(1)}_{IR}(T_{dip})\vert 0^{A}_k,\,1^{B}_k\rangle=\frac{\vert 0^{A}_k,\, 1^{B}_k\rangle -i \,\vert 1^{A}_k,\, 0^{B}_k\rangle }{\sqrt{2}}.
\end{equation}
In the next Section we discuss the propagation of the two-polariton state in Eq. \ref{eq:initial_state} through a generic QPIC. Interestingly, it is possible to map the polariton-polariton dynamics into that of a two-level atom driven by external electromagnetic pulses. 

\begin{figure}
    \centering
    \includegraphics[scale=0.7]{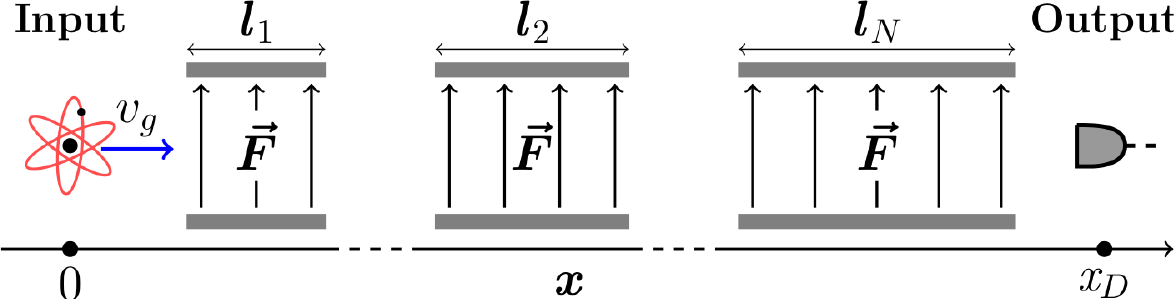}
    \caption{A physical system described by the propagation of the two-polariton state defined in Eq. \ref{eq:initial_state} through the generic QPIC represented in Fig. \ref{fig:generic_device}. A two-level atom starts from the left with velocity equal to $v_g$, i.e. the polariton group velocity, and propagates towards the detection region (output side), passing through regions characterized by the presence or the absence of an external transverse field $\vec{F}$.}
    \label{fig:equivalent_two_atom_system}
\end{figure}

\subsection{Dynamics of two identical polaritons}\label{sec:atom_polariton_correspondence}
Starting from a double excitation subspace (i.e., an initial configuration with a total number of two polaritons in the system), the quantum state evolves in time as the following superposition  
\begin{equation}
\vert \psi(t)\rangle = \alpha(t) \vert 1^{A}_k,\, 1_k^{B}\rangle+\beta(t) \vert 2_k^{A},\, 0^{B}_k\rangle+\gamma(t) \vert 0_k^{A},\, 2^{B}_k\rangle \, ,
\end{equation}
where we introduced the states $\vert 2_k^{A},\, 0^{B}_k\rangle=(\hat{a}_k^{\dagger})^2\vert \Omega \rangle /\sqrt{2}$ and $ \vert 0_k^{A},\, 2^{B}_k\rangle=(\hat{b}_k^{\dagger})^2\vert \Omega \rangle/\sqrt{2}$. Therefore, in the general case, the time evolution of any two-polariton input state can be effectively described by means of a three level system, and the coefficients $\alpha(t)$, $\beta(t)$, and $\gamma(t)$ are only related by the normalization condition on $\vert \psi \rangle$.
However, by considering that the Hamiltonian is symmetric under the exchange of $\hat{a}_k$ and $\hat{b}_k$, the input state in Eq. \ref{eq:initial_state} evolves as:
\begin{equation}
\vert \psi(t)\rangle = \alpha(t) \vert 1^{A}_k,\, 1_k^{B}\rangle+\beta(t) \left(\vert 2_k^{A},\, 0^{B}_k\rangle+\vert 0_k^{A},\, 2^{B}_k\rangle\right) \, .
\end{equation}
Therefore, by defining
\begin{equation}
    \vert g \rangle = \vert 1^{A}_k,\, 1_k^{B}\rangle,\quad \vert e \rangle= \frac{\vert 2_k^{A},\, 0^{B}_k\rangle+\vert 0_k^{A},\, 2^{B}_k\rangle}{\sqrt{2}},
\end{equation}
the Hamiltonian model in Eq. \ref{eq:model_Hamiltonian} reduces to 
\begin{equation}\label{eq:two_level_hamiltonian}
    \mathcal{\hat{H}}=(2\omega_k+U_k) \mathbbm{1}+2J(x)\sigma_x -U_k \sigma_z \, ,
\end{equation}
where $\mathbbm{1}$, $\sigma_x$, and $\sigma_z$ are the identity and the Pauli-X and Z matrices, respectively. In particular, in our representation $\sigma_z\vert\, g\,\rangle= \vert\, g\,\rangle$ and $\sigma_z\vert\, e\,\rangle= -\vert\, e\,\rangle$.\\
This model equivalently describes a two-level atom propagating in free space with transition frequency/detuning given by $2U_k$, and effectively interacting with a transverse field, $\vec{F}$, producing a Rabi frequency $2J(x)$, and $ \vert g \,\rangle$  ($\vert e \,\rangle$) corresponds to the ground (excited) state of such fictitious atom. In other words, the propagation of the state in Eq. \ref{eq:initial_state} through the QPIC reported in Fig. \ref{fig:generic_device}, is equivalent to the physical situation pictured in Fig. \ref{fig:equivalent_two_atom_system}. On the left (input side), there is an atom initially prepared in the $\vert\,g\,\rangle$ state. Such a particle is then sent at speed $v_g$ through a sequence of regions having different properties, i.e. with or without the external transverse field $\vec{F}$, and its properties are read on the rightmost side of the apparatus (output region). In particular, we assume that in between the regions represented in Fig. \ref{fig:equivalent_two_atom_system} and characterized by a length $l_j$, there are regions where the atom can propagate freely, i.e. $\vec{F}=\vec{0}$. Similarly to what happens in the one particle sector, the faster the atom moves (i.e., the larger $v_g$), the smaller is the time spent by polaritons in each hopping regions, and thus the smaller is the probability of observing a polariton jumping from a waveguide to the adjacent one. On the other hand, in the two-particle subspace of the Hamiltonian \ref{eq:model_Hamiltonian}, the terms depending on $U_k$ play a non negligible role. In particular, following Eq. \ref{eq:two_level_hamiltonian}, and by virtue of the properties of the Pauli matrices, the propagation within the polariton interferometer depicted in Fig.~\ref{fig:generic_device} is completely determined by composing the proper number of times two unitary operators
\begin{equation}\label{eq:beam_splitter_operator}
\begin{split}
&U^{(2)}_{IR}(t)=e^{-i\phi t}\left[\cos(\sqrt{U_k^2+4J^2} t)\mathbbm{1}+\right.\\
&\left.+i\frac{\sin(\sqrt{U_k^2+4J^2}t)}{\sqrt{U_k^2+4J^2}} \left(-2J\sigma_x+U_k\sigma_z\right)\right] \, ,
\end{split}
\end{equation}
and 
\begin{equation}\label{eq:free_propagation_operator}
    U^{(2)}_{FP}(t)=e^{-i\phi t}\left[\cos(U_k t)\mathbbm{1}+i\sin(U_k t)\sigma_z\right] \, ,
\end{equation}
where $\phi=2\omega_k +U_k$ and the subscript FP stands for "free propagation". The operator in Eq. \ref{eq:beam_splitter_operator} describes the evolution of an initially symmetric state along an interaction length $l=v_g t$, while the one in Eq. \ref{eq:free_propagation_operator} describes the evolution of the two-polariton system in the free propagation region between two interaction regions. In particular, notice that $U^{(2)}_{FP}(t)$ introduces a phase difference between the states $\vert g \rangle$ and $\vert e \rangle$, which is proportional to $2U_k$. 

\section{Results} \label{sec:lossless_devices}

In this Section we consider two paradigmatic cases of the generic QPIC {showing how these devices can be used to implement nonlinear quantum gates and for precisely measuring the interaction strength between two single polaritons.} 
A generalized version of the solution corresponding to the case of ideally lossless polariton propagation is reported in SI. In particular, since single polariton states are not affected by $U_k$, we consider only the evolution of the state defined in Eq. \ref{eq:initial_state}. 

\subsection{Single hopping region: \\ {nonlinear HOM effect and $\sqrt{SWAP'}$ gate}}\label{subsec:HOM_lossless}

\begin{figure}[t]
    \centering
    \includegraphics[scale=0.75]{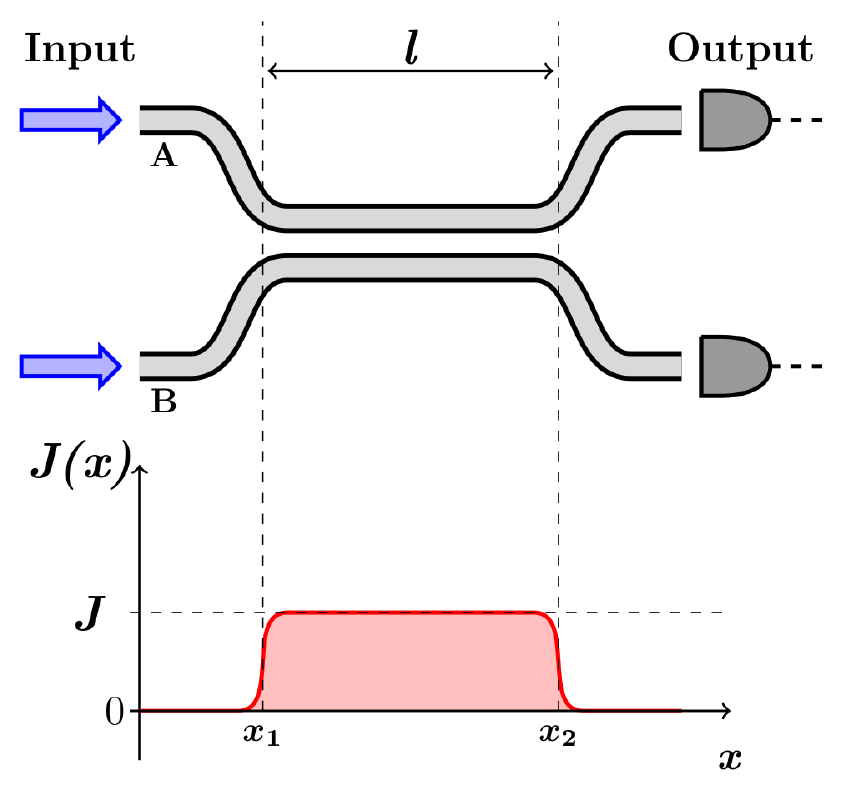}
    \caption{Schematic representation of the elementary QPIC building block, in which a pair of co-propagating waveguides is coupled along a spatial region of length $l$. In the bottom part of the panel we sketch the behavior of the coupling constant, $J(x)$, which is non-zero and equal to a constant value ($J$) only in the region $x_1<x<x_2$. }
    \label{fig:HOM_setup}
\end{figure}
 The most elementary building block of a QPIC, namely a single region of evanescent coupling between waveguides $A$ and $B$, characterized by the length $l$, is depicted in Fig.~\ref{fig:HOM_setup}. This configuration is typically exploited in conventional PCs to realize a beam splitter. Here, the main novelty is the presence of the polariton number-dependent nonlinear interaction within the single waveguide channel, which is going to alter the beam splitting condition, as well as influencing the many particle statistics in each channel.

In order to characterize the quantum behavior of the device, we will consider the auto- and cross-correlation functions at the output ports, defined as
\begin{equation}\label{eq:G_AA}
G_{AA}(t,\,U_k/J)= \langle \psi(t)\vert \hat{a}^\dagger_k\,\hat{a}^\dagger_k\,\hat{a}_k\,\hat{a}_k\vert  \psi(t)\rangle \, ,
\end{equation}
and
\begin{equation}\label{eq:G_AB}
G_{AB}(t,\,U_k/J)= \langle \psi(t)\vert \hat{a}^\dagger_k\,\hat{b}^\dagger_k\,\hat{b}_k\,\hat{a}_k\vert  \psi(t)\rangle \, ,
\end{equation}
where $t$ is the polariton arrival instant at the rightmost end of the device. Here, polaritons are converted in photons. Therefore, the former amplitude is the probability of having two photons emerging in pairs from the waveguide $A$ (notice that, by symmetry, $G_{AA}(t,\,U_k/J)=G_{BB}(t,\,U_k/J)$). The latter amplitude, $G_{AB}(t,\,U_k/J)$, accounts for events related to photons simultaneously emerging (and being detected) from different waveguides. In particular, notice that both quantities can be readily measured in a realistic quantum optical experiment.
Following Eq. \ref{eq:beam_splitter_operator}, after some algebra, analytic expressions can be obtained, which read:
\begin{equation}\label{eq:GAA_HOM}
G_{AA}(t,\,U_k/J)=\frac{2J^2}{U_k^2+4J^2}\sin^2(\sqrt{U_k^2+4J^2}t)
\end{equation}
and 
\begin{equation}\label{eq:GAB_HOM}
    G_{AB}(t,\,U_k/J)=1-2 G_{AA}(t,\,U_k/J) \, .
\end{equation}

\begin{figure}[t]
    \centering
    \includegraphics[scale=0.5]{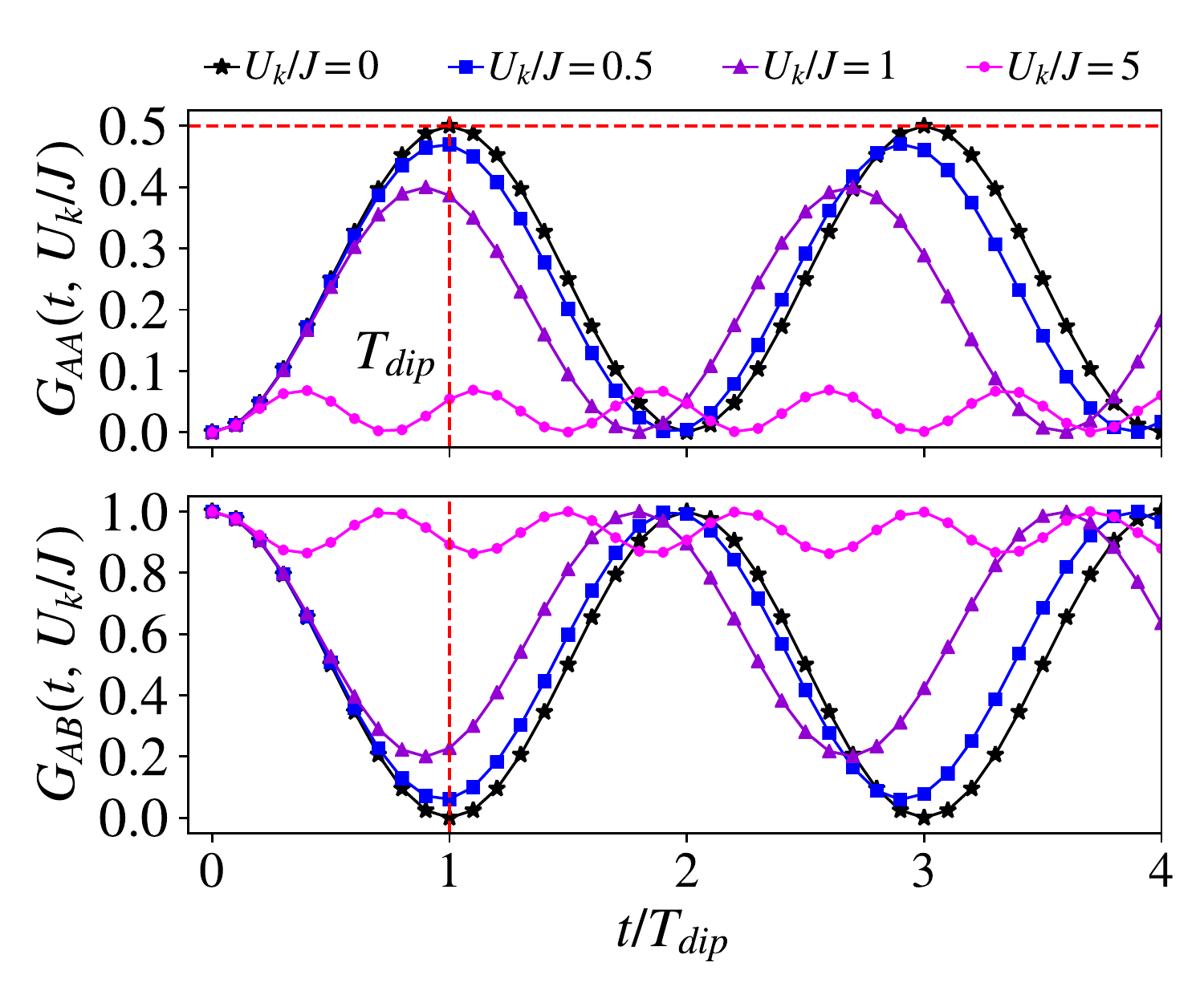}
    \caption{Dependence of the auto- and cross-correlation functions, $G_{AA}(t,\,U_k/J)$ and $G_{AB}(t,\,U_k/J)$, on the time spent within the hopping region of length $l$,  for different values of the ratio $U_k/J$. The hopping time, $t$, is given in units of $T_{dip}=\pi/(4J)$, which corresponds to HOM condition in the absence of nonlinearities in the propagating channels, $U_k=0$. The vertical dashed lines in both panels identify the HOM condition ($t/T_{dip}=1$).}
    \label{fig:correlations_HOM}
\end{figure}

\begin{figure}[t]
    \centering
    \includegraphics[scale=0.5]{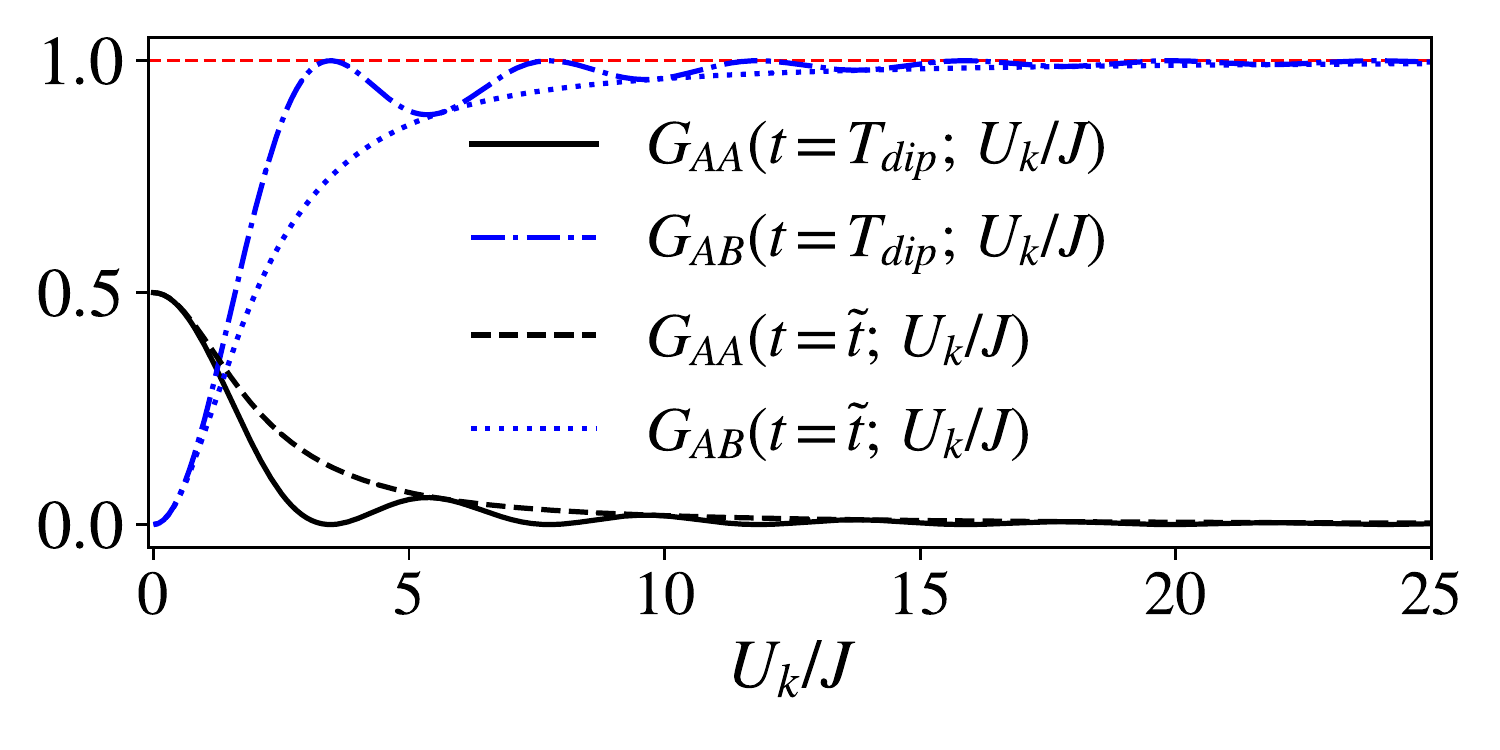}
    \caption{Dependence of correlation functions at the HOM condition as a function of the polariton interaction strength, $G_{AA}(t=T_{dip},\,U_k/J)$ and $G_{AB}(t=T_{dip},\,U_k/J)$. Dotted and dashed curves describe the behavior at increasing $U_k/J$ of the maximum and minimum values reached by the same correlation functions for $t= \tilde{t}$.}
    \label{fig:maxima_HOM}
\end{figure}

The dependence of $G_{AA}$ and $G_{AB}$ on $U_k/J$ is shown in Fig.~\ref{fig:correlations_HOM}. In particular, here we plot the dependence of these correlation functions on the hopping time spent by the two polariton wave packets in the region of length $l$. This time is compared to the parameter $T_{dip}$, which corresponds to the time for which the device would behave as a regular 50:50 beam splitter in the absence of polariton interactions, i.e. for $U_k/J=0$. In fact, this can be interpreted as the time for which the interferometer results in the well known Hong-Ou-Mandel (HOM) effect in conventional PCs.\cite{hong_measurement_1987,Politi2008,Heeres2013plasmonHOM,Spagnolo2013bosonbunching,luo_nonlinear_2019} Such an established phenomenon is related to the bosonic nature of two indistinguishable particles simultaneously impinging on a 50:50 beam splitter from two different input ports, and it is characterized by a dip in the probability of observing two particles emerging simultaneously from the two different output ports, ideally going to zero. In our notation, this corresponds to the condition for which $G_{AB}(T_{dip},\,0)=0$. At the same time, this condition also implies that the probability of observing two bosons emerging in pairs from the same waveguide is maximized (boson bunching), i.e. $G_{AA}(T_{dip},\,0)=G_{BB}(T_{dip},\,0)= 0.5$. This is exactly what is shown in the plots of Fig.~\ref{fig:correlations_HOM} for $U_k/J=0$ and $t/T_{dip}=1$, with an obvious periodicity as a function of $t/T_{dip}$. However, while the HOM effect has been mostly explored in linear interferometers, so far, our framework allows one to easily study the effect of polariton nonlinearities. 
In particular, the HOM dip tends to be suppressed on increasing $U_k/J$, i.e. strong polariton correlations inhibit the quantum interference at the beam splitting condition. More in detail, as suggested by the results in Eqs.~\ref{eq:GAA_HOM} and \ref{eq:GAB_HOM}, for $U_k/J \gg 1$ (i.e., what we define the strongly correlated polariton regime) the probability of observing photons emerging in pairs from the same waveguide becomes negligible, while photons are mostly expected to be simultaneously detected from the two different output channels, i.e. 
\begin{equation}\label{eq:strong_limit}
 G_{AA}(t,\,U_k/J)\to 0,\quad   G_{AB}(t,\,U_k/J)\to 1  
\end{equation}
The detailed convergence to this regime as a function of the relevant scale $U_k/J$ is explicitly shown in Fig.~\ref{fig:maxima_HOM}. Here we plot the behavior of the two amplitudes defined in Eqs. \ref{eq:GAA_HOM} and \ref{eq:GAB_HOM} for $t/T_{dip}=1$, and the maximum and minimum value they reach respectively while varying the ratio $t/T_{dip}$, i.e. corresponding to an interaction time $\tilde{t}\equiv \pi/(2\sqrt{U^2_k+4J^2}) $.

Essentially, this behavior can be interpreted as follows: on increasing the on-site interaction within each propagating channel, the bunching probability is suppressed. In quantum photonics, this is reminiscent of the polariton blockade effect \cite{verger_blockade_2006,Gerace-NV-nmat-2019}, recently evidenced in confined geometries \cite{Matutano-emergence-2019,Delteil-blockade-2019}. Evidently, if such a regime could be accessed in a propagating geometry, the device would implement a kind of nonlinear beam splitter in which bosonic coalescence is destroyed in favor of ``fermionization'' induced by strong interactions \cite{Carusotto2009fermionization,Gerace2009nphys}. \\
{Noticeably, in the limit $U_k\gg J$ such a device would also implement a $\sqrt{\mbox{SWAP}'}$ quantum gate \cite{Franson2004}, when a single rail qubit encoding is assumed: each waveguiding channel is in logic $\vert 0\rangle$ state if no polariton is present, and logic $\vert 1\rangle$ with a single propagating polariton state. Indeed, it is easy to verify that for $t=T_{dip}$, when using $\{\vert 0^A_{k},\, 0^B_{k}\rangle,\,\vert 1^A_{k},\, 0^B_{k}\rangle,\,\vert 0^A_{k},\, 1^B_{k}\rangle,\,\vert 1^A_{k},\, 1^B_{k}\rangle\}$ as the two-qubit basis set, the QPIC in Fig. \ref{fig:HOM_setup} is described by the following matrix operation 
\begin{equation}\label{eq:sqrt_swap_gate}
\sqrt{SWAP'}=\left(
    \begin{matrix}
    1 & 0 & 0 & 0 \\
    0 & \frac{1}{\sqrt{2}} & \frac{-i}{\sqrt{2}} & 0 \\
    0 & \frac{-i}{\sqrt{2}} & \frac{1}{\sqrt{2}} & 0 \\
    0& 0& 0& 1\\
    \end{matrix}\right) \, ,
\end{equation} 
which allows to implement universal quantum computing when combined with arbitrary single qubit rotations \cite{Nielsen_Chuang}. }

More practically, even being limited to the realm of weak nonlinearities that do not allow to access such strong correlation regime, this configuration may still allow to directly assess interaction-induced deviations from the HOM condition. However, as it will be discussed in Sec.~\ref{sec:parameters}, polariton interaction energies have been reported in the few $\mu$eV range, which appear to be too small to be detected on the scale of the plot in Fig.~\ref{fig:maxima_HOM} ($J$ is typically in the meV order of magnitude). 
The scheme in Fig. \ref{fig:HOM_setup}, on the other hand, is just the simplest building block for QPICs. By exploiting a slightly more complex scheme it is possible to design polariton circuits showing enhanced sensitivity to the critical parameter $U_k / J$ that, in turn, can be exploited to implement quantum gates even in the presence of small nonlinearities, which will be shown in the following.

\begin{figure}[t]
    \centering
    \includegraphics[scale=0.5]{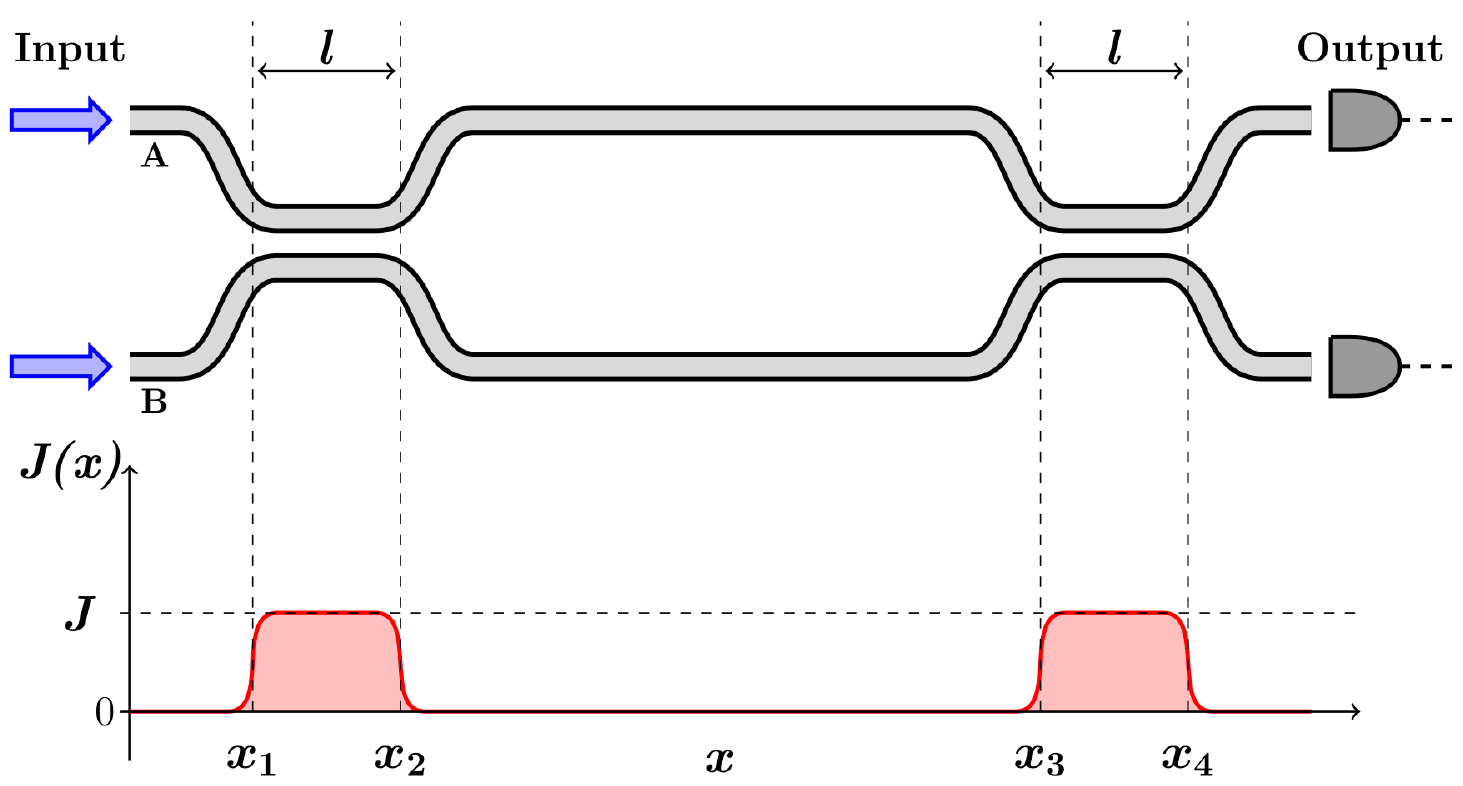}
    \caption{Schematic representation of an analog Ramsey interferometer in QPIC: two hopping regions of length $l$ are separated by a free propagation region. In the bottom panel, the spatial dependence the hopping parameter along the device length is sketched. As in the previous case, we consider $J(x)=J$ in the two interaction regions and zero elsewhere.}
    \label{fig:Ramsey_setup}
\end{figure}

\subsection{Two hopping regions: polariton Ramsey interferometer for enhanced sensitivity } 
Motivated by the results shown in the previous section, and by the atom-polariton correspondence discussed before in Sec. \ref{sec:atom_polariton_correspondence}, here we explore the dynamics and the resulting quantum correlations at the output of the device depicted in Fig.~\ref{fig:Ramsey_setup}. In this case, the hopping is split into two separate regions having the same length, $l$, and separated by a free propagation region within each waveguide. Similarly to the previous Section, we define the time spent in each interaction region, $t$, and the time spent in the region $x_2 < x <x_3$, $T$, as the relevant parameters. Given these definitions, the action of the device in Fig.~\ref{fig:Ramsey_setup} on the two polariton state defined in Eq.~\ref{eq:initial_state} is described by the unitary operator
\begin{equation}
    U(t,\,T)\equiv U_{IR}(t) U_{FP}(T) U_{IR}(t) \,.
\end{equation}

\begin{figure}[t]
    \centering
    \includegraphics[scale=0.55]{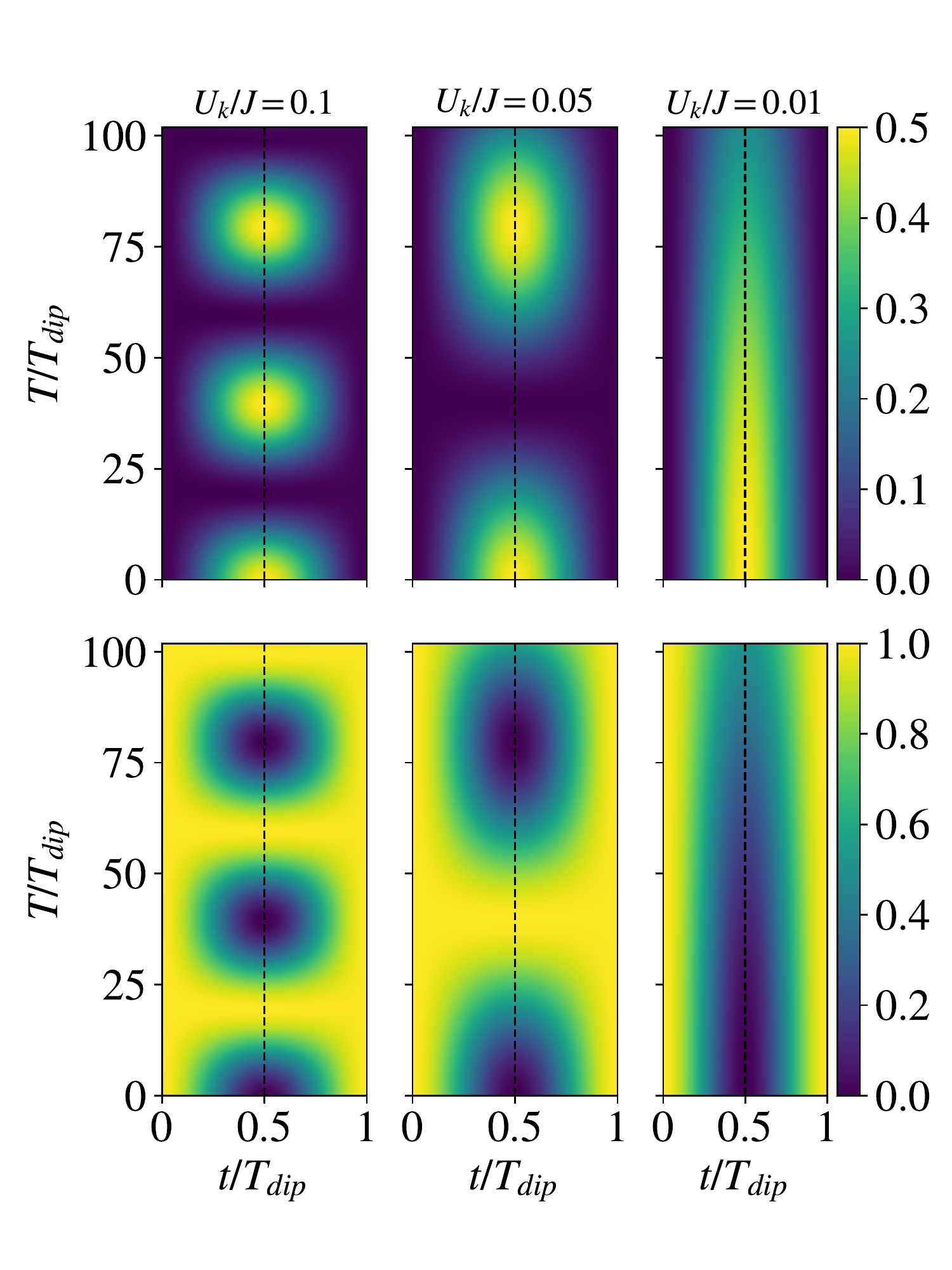}
    \caption{Behavior of correlation functions in the analog Ramsey setup. The three top (bottom) panels describe  $G_{AA}(t,\,T;\,U_k/J)$ ($G_{AB}(t,\,T;\,U_k/J)$) for different values of $U_k/J$, as a function of both $t/T_{dip}$ and $T/T_{dip}$. The particular value of $U_k/J$ considered is specified on top of each column. Dashed vertical lines have been added to each panel to guide the eye in evidencing the condition $t/T_{dip}=0.5$.}
    \label{fig:Global_behavior_Correlation_Ramsey}
\end{figure}

In particular, a direct analytic computation shows that the probability of having two photons being detected from the same output channel after the second hopping region reads
\begin{equation}\label{eq:G_AA_Ramsey}
\begin{split}
    &G_{AA}(t,\,T;\,U_k/J)=
    \frac{2J^2}{(U_k^2+4J^2)}\sin^2(2\theta t)\cos^2(U_k T)\\
    &-\frac{2J^2 U_k}{(U_k^2+4J^2)^{3/2}}\sin^2(\theta t)\sin(2\theta t)\sin(2U_kT)\\
    &+\frac{8J^2U_k^2}{(U_k^2+4J^2)^2}\sin^4(\theta t)\sin^2(U_k T),\\
\end{split}
\end{equation}
with $\theta=\sqrt{U_k^2+4J^2}$. Similarly to the previous case, we have $G_{BB}=G_{AA}$ and 
\begin{equation}\label{eq:G_AB_Ramsey}
    G_{AB}(t,\,T;\,U_k/J)=1-2 G_{AA}(t,\,T;\,U_k/J) \, .
\end{equation}
Here, it is worth highlighting the explicit dependence of the correlation functions on $t$, for which the degree of correlation of the output photons can be modulated by either changing the interaction length, $l$, or the separation between the two hopping regions, thus the time $T$. 

Globally, the  behavior of $G_{AA}(t,\,T;\,U_k/J)$ and $G_{AB}(t,\,T;\,U_k/J)$ for different values of $U_k/J$, and as a function of both $t/T_{dip}$ and $T/T_{dip}$, is given in the plots of Fig.~\ref{fig:Global_behavior_Correlation_Ramsey}. Correlation functions display a periodic dependence on $T/T_{dip}$. In particular, for all the values of $U_k/J$ considered, the probability of having photons emerging simultaneously from the same waveguide, quantified by $G_{AA} (t,\,T;\,U_k/J)$, is maximized close to the condition $t/T_{tip}=0.5$. The first maximum is located in all the cases at $T/T_{dip}=0$, i.e. the condition for which the analog  setup reduces to the one characterized in the previous section ($2t=T_{dip}$ and $T=0$). The following peaks of $G_{AA} (t,\,T;\,U_k/J)$ as a function of $T/T_{dip}$  explicitly depend on $U_k/J$. Due to their complementarity, $G_{AB} (t,\,T;\,U_k/J)$ is minimized whenever $G_{AA} (t,\,T;\,U_k/J)$ is maximized, as it is evident from the bottom panels of Fig.~\ref{fig:Global_behavior_Correlation_Ramsey}. 

\begin{figure}[t]
    \centering
    \includegraphics[scale=0.5]{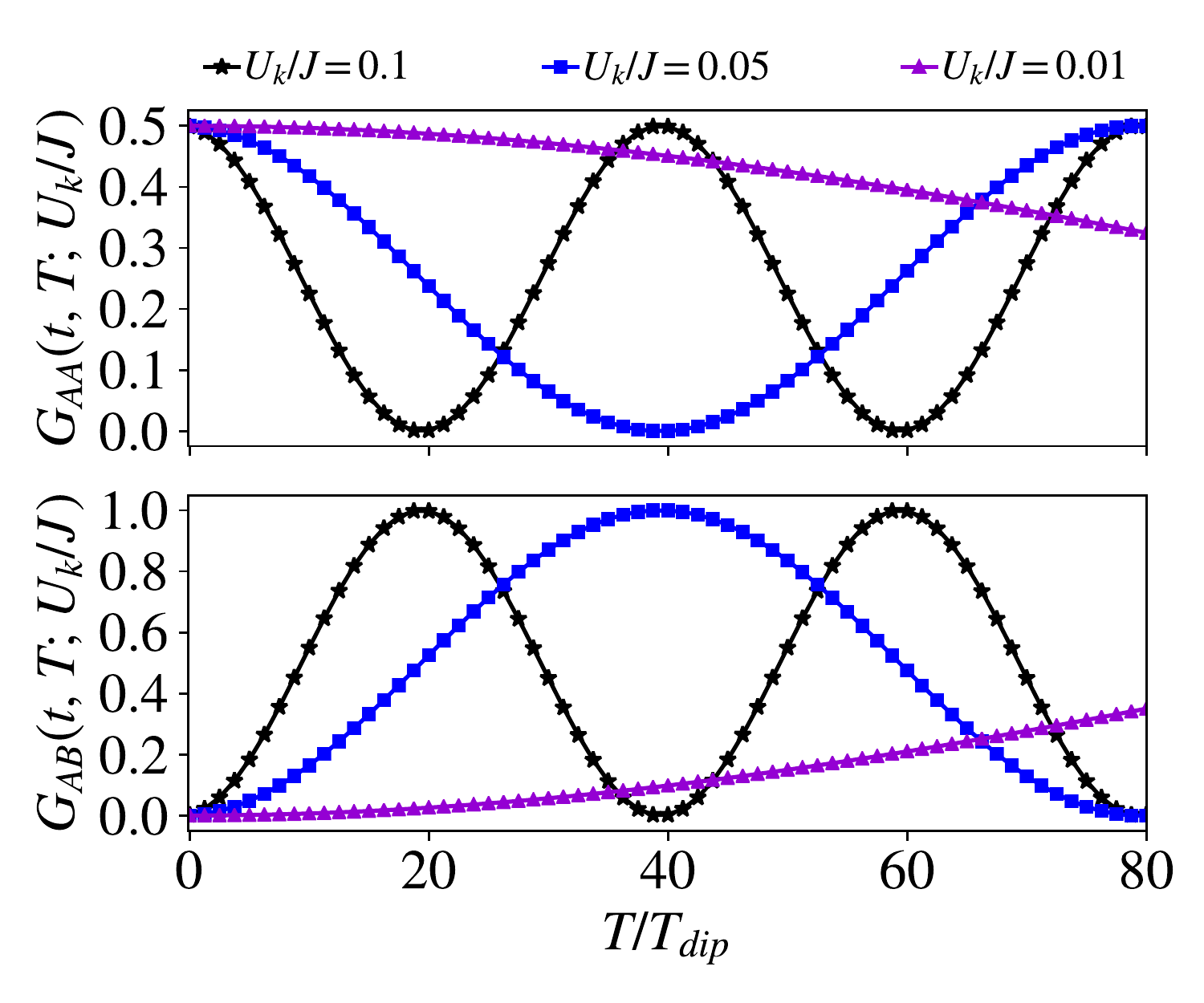}
    \caption{Dependence of $G_{AA}(t,\,T;\,U_k/J)$ and $G_{AB}(t,\,T;\,U_k/J)$ on $T/T_{dip}$, for fixed $t/T_{dip}=0.5$, and for the different values of $U_k/J$ specified on top of each panel.}
    \label{fig:Ramsey_Correlations_Cut}
\end{figure}

In addition, it is worth stressing that these results suggest that the geometry assumed for the device in Fig.~\ref{fig:Ramsey_setup} is considerably more sensitive to small values of the ratio $U_k/J$, if compared to the elementary interferometer considered in the previous section. In order to better clarify this point, let us consider a selection of results plotted in Fig.~\ref{fig:Ramsey_Correlations_Cut}, where we show both $G_{AA} (t,\,T;\,U_k/J)$ and 
$G_{AB} (t,\,T;\,U_k/J)$ as a function of $T/T_{dip}$, at fixed $t/T_{dip}=0.5$ (i.e., following the vertical dashed lines in Fig.~\ref{fig:Global_behavior_Correlation_Ramsey}).
In analogy to the Ramsey interferometer, here significant variations in the degree of correlation between the photons emerging at the output of the device are obtained by changing the ratio $T/T_{dip}$. The latter essentially coincides with varying the distance $L=x_3-x_2$ between the two interaction regions for a given choice of the hopping distance ($d$) and the group velocity of the propagating mode (see, e.g., the relative section in SI).
We propose, in particular, to employ this QPIC to extract information about the value $U_k$ (knowing the value of $J$), by considering the output correlations obtained with devices having an interaction length equivalent to $t/T_{dip}\approx 0.5$ and different values of $T/T_{dip}$. Notice that $t=0.5 T_{dip}$ corresponds to an interfering region that is only half of the length corresponding to the beamsplitting condition in the linear ($U_k=0$) regime, which by our definition is given for $t=T_{dip}$. The reason for the increased sensitivity to the small nonlinear shift can then be traced back to the same reason giving the high sensitivity to small energy shifts in the conventional Ramsey interferometer, given the formal analogy reported above \cite{RamseyBook}.

Finally, such enhanced sensitivity can be extremely beneficial for the realization of quantum gates, such as SWAP-type gates, even in presence of small nonlinearities. Indeed, according to the results reported in Fig. \ref{fig:Ramsey_Correlations_Cut}, it can immediately be recognized that for $t=T_{dip}/2$ and  $T/T_{dip}\approx 2J/U$, $G_{AA}\approx 0 $ and $G_{AB}\approx 1$. 
In other words, for such values of propagation times the QPIC in Fig. \ref{fig:Ramsey_setup} exactly behaves like the $\sqrt{\mbox{SWAP}'}$ gate defined in Eq.~\ref{eq:sqrt_swap_gate} on the basis set $\{\vert 0^A_{k},\, 0^B_{k}\rangle,\,\vert 1^A_{k},\, 0^B_{k}\rangle,\,\vert 0^A_{k},\, 1^B_{k}\rangle,\,\vert 1^A_{k},\, 1^B_{k}\rangle\}$, even for weak values of $U_k$. The advantages of this scheme are noteworthy, since no post-selection is required to implement the gate, and no errors due to state occupancy beyond the computational basis can occur. This result holds great promise to realize prospective devices for photonic quantum computing with deterministic gates. 

\section{Discussion:\\experimental implementation}\label{sec:parameters}

Here we assess the relevance of the previous theoretical analysis in view of potential applications and experiments in quantum polaritonics. As already mentioned in the Introduction, the polariton-polariton interaction energy has proven to be a quantity that is hard to be probed directly in experiment, although a few reliable estimates exist in the literature \cite{Matutano-emergence-2019,Delteil-blockade-2019}. Moreover, such nonlinearities have been shown to be strongly enhanced in suitably engineered nanostructures and exploiting the application of an external electric field \cite{Rosenberg2018}, although the precise enhancement has later been reduced \cite{Suarez-Forero2021}. Following recent estimates, we assume a polariton nonlinear interaction energy as large as $U=15$ $\mu$eV in standard III-V semiconductor technology \cite{Delteil-blockade-2019}, when the polariton field is confined in a spot on the order of 1 $\mu$m$^2$. In a propagating geometry, even if a polariton wave packet is somehow localized in space, this value might be assumed of the same order of magnitude. Then, the latter could be further enhanced by a factor of 5-10 due to electric field tuning, depending on the different systems reported in the literature \cite{Rosenberg2018,Suarez-Forero2021}, or even exploiting different mechanisms exploiting coupling to the biexciton state \cite{Carusotto_EPL_2010}. \\
In this Section we provide some estimates on the values of $U_k$ that can be detected by a quantum polaritonic  interferometer in realistic scenarios. First, we consider an experimental setup with a single interaction region. A similar device has been experimentally realized, e.g., in Ref.~\cite{Beierlein2021}. While the focus of that work is different, and none of the theoretical results reported here can be benchmarked, we can still use some of the experimentally determined figures of merit to derive realistic model parameters. Referring to the quoted work, by shining a laser on one of the two waveguides, e.g., the ``A" channel in our definition, the authors first created a coherent superposition of many-polariton states, and then characterized its propagation along the device. In particular, they studied how tunneling is affected by the gap between the two channels (i.e., the spatial separation $d$ in our case, see Fig. \ref{fig:generic_device}), and provided a realistic estimate for the tunneling energy, $J$. It is seen that values on the order of $J=0.5$ meV are experimentally realistic. Hence, the relevant ratio $U_k /J$ would range from $10^{-2}$ to $10^{-1}$ in state-of-art devices. These values appear too small to provide a significant deviation from the device operated in a linear regime, as shown in Figs. \ref{fig:correlations_HOM} and \ref{fig:maxima_HOM}. On the other hand, such values could be easily detected from correlation measurements in the analog Ramsey setup of Fig.~\ref{fig:Ramsey_setup}, according to the results reported in Figs.~\ref{fig:Global_behavior_Correlation_Ramsey} and \ref{fig:Ramsey_Correlations_Cut}, by increasing $T$.
\\
{More explicitly, we consider the case of a cross-correlation value $G_{AB}(t=T_{dip}/2,\,T;U_k/J)=0.2$, that is the physical situation in which $20\%$ of the events correspond to double clicks, i.e., two photons emerging simultaneously from the different waveguides, as a targeted detectable level of coincidences above the noise. Depending on the specific value of the polariton nonlinearity, such a photon counting will occur in correspondence to  different values of $T$. Let us denote the first time at which such an event occurs as $T^*$. Since $T^* + T_{dip}$ corresponds to the total time spent by polaritons within the device, multiplying this quantity by $v_g$ gives the total Ramsey interferometer length leading to the condition mentioned above (i.e., $20\%$ coincidence events), $l_{tot}$. A few realistic values for $l_{tot}$ can be obtained by using $J=0.5$ meV (which corresponds to $T_{dip}\approx 1$ ps) and a group velocity in the range 4 to 5 $\mu$m/ps, which is compatible with the polariton group velocity  characterized, e.g., in Ref. \cite{Beierlein2021} 
(see also SI for further details). A few numerical examples are reported in Table~\ref{tab:total_length}, obtained for the ratios of normalized parameter $U_k/J$ considered throughout the previous Sections.}

\begin{table}[]
    \centering
    
    \begin{tabular}{|c|c|c|c|}\hline
   $U_k/J$ & $U_k$  &$T^{*}$ & $l_{tot}=v_g (T^*+T_{dip})$ \\ 
    $0.1$  & $50$ $\mu$eV & $5.3$ ps & $25 - 32$  $\mu$m \\
    $0.05$  & $25$ $\mu$eV & $11.1$ ps & $48 - 61$ $\mu$m \\
    $0.01$  & $5$ $\mu$eV & $58.6$ ps & $240 - 300$ $\mu$m \\
    \hline
    \end{tabular}
    \caption{Numerical estimates for total length of the polariton Ramsey interferometer for which $G_{AB}(t=T_{dip}/2,\,T^*;U_k/J)= 0.2$, i.e., $20\%$ of the events corresponding to double clicks. We used $J=0.5$ meV (corresponding to a $T_{dip}\sim 1$ ps) and a group velocity in the range $4 - 5$ $\mu$m/ps.}
    \label{tab:total_length}
\end{table}

\begin{figure}
    \centering
    \includegraphics[scale=0.55]{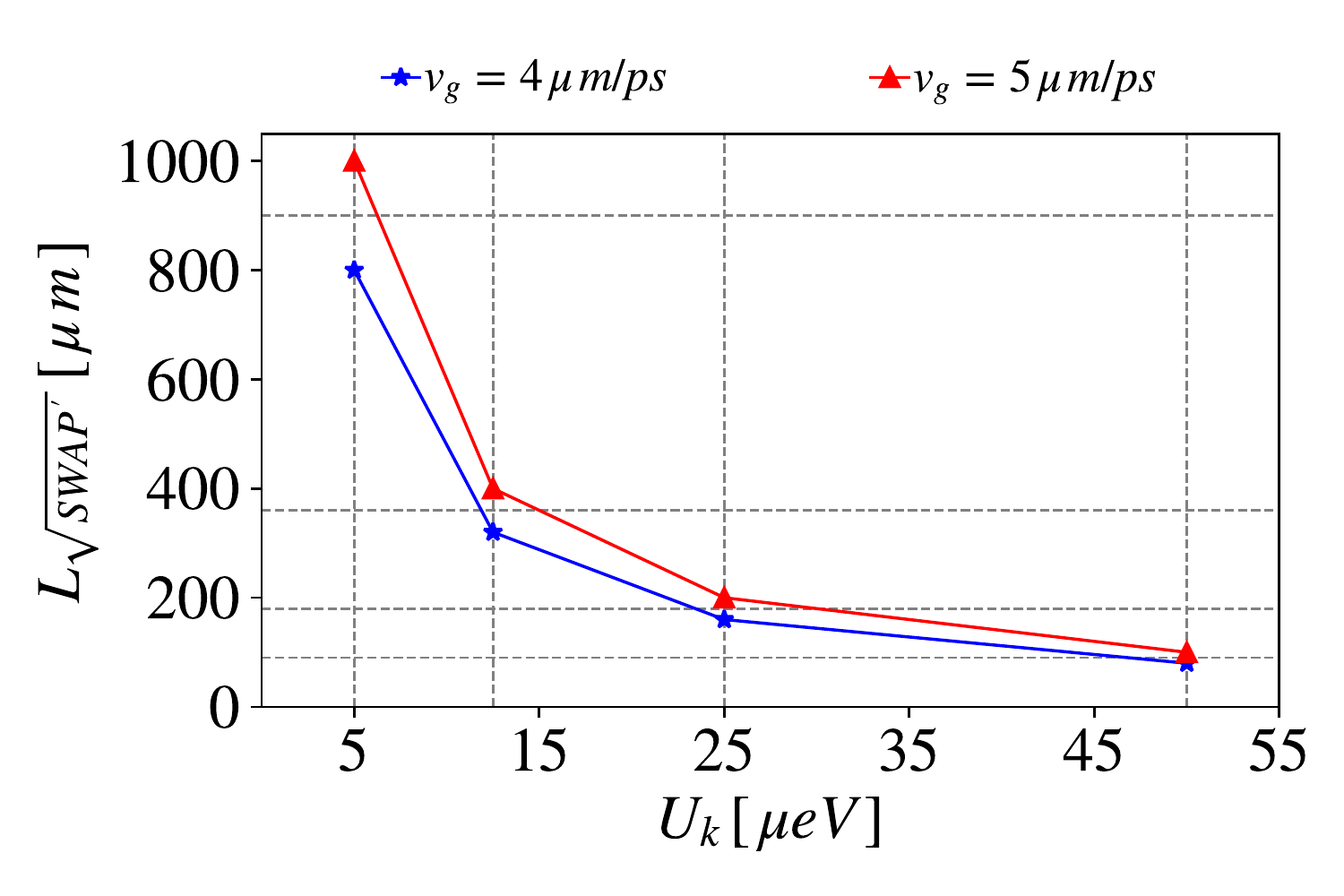}
    \caption{Behavior of the free propagation length $L_{\sqrt{SWAP'}}$ as a function of the nonlinearity $U_k$, for two different values of the group velocity. The dashed horizontal and vertical lines have been added to guide the eye.}
    \label{fig:Ramsey_SWAP}
\end{figure}

Finally, concerning the specific application of the Ramsey-like interferometric configuration in view of implementing a SWAP-type quantum gate, we explicitly report some realistic device parameters in Fig.~\ref{fig:Ramsey_SWAP}. In particular, we address the dependence of free propagation region length for the specific operation in Eq.~\ref{eq:sqrt_swap_gate}, defined as $L_{\sqrt{SWAP'}}$, as a function of the nonlinearity $U_k$, for some values in the supposedly realistic range $5\sim 50\,\mu eV$. This length is numerically calculated by following the condition for which $G_{AA}=0.0$ (or $G_{AB}=1.0$) in the plot of Fig.~\ref{fig:Ramsey_Correlations_Cut}, and converting the time $T$ into a length $v_g T$.
Similarly to the previous discussion, we consider two values for the group velocity (as explicitly given in the Figure). Interestingly, considering that $U_k \simeq 25$ $\mu$eV can be realistically expected on the basis of experimental results reported in the literature, a free propagation length of about 200 $\mu$m should be realized. The latter corresponds to about $T\sim 40-50$ ps, which is fully within the typical polariton lifetime of $\sim 100$ ps in these systems. Combining this propagation length with the two interfering regions, the overall length of the device should allow to perform a realistic proof-principle demonstration of this quantum gate.

\section{Conclusion}\label{sec:conclusions}

We have introduced a novel paradigm for the development of quantum technologies that extends current performances of photonic integrated circuits, by exploiting effective interactions between polaritons. We have shown that they allow to build nonlinear quantum devices and deterministic quantum logic gates. As a first step, we have analyzed how polariton interactions can induce a significant deviation from the typical Hong-Ou-Mandel behavior in an integrated beam splitter. Starting from this result we then explicitly designed an integrated polariton interferometer, which allows to straightforwardly implement a deterministic $\sqrt{\mbox{SWAP}'}$ gate. By considering realistic experimental conditions and parameters, the proposed QPIC represents a viable route for prospective photonic-based quantum information processing.

Noticeably, the interferometric nature of  our QPICs make these devices very well suited also for metrological and sensing purposes. As an targeted example, we have shown how an analog integrated Ramsey interferometer could be used for measuring polariton interaction energies at two particle level.

In conclusion, we believe these results will motivate the realization of complex polariton interferometers, which will set the basis of future experiments exploiting the relatively strong nonlinearities of these systems for metrological applications, sensing, quantum simulations and quantum information processing.

\acknowledgments
We acknowledge financial support from the Italian Ministry of Research (MIUR) through the PRIN 2017 project ``Interacting photons in polariton circuits'' (INPhoPOL). Useful discussions with  V. Ardizzone, D. Ballarini, A. Gianfrate, E. Maggiolini, D. Su\'{a}rez-Forero are gratefully acknowledged.  

\newpage

\section*{Supplementary Information}

\subsection{Derivation of the polariton Hamiltonian}
\label{app:polaritons-derivation}

In this section we derive the standard model used to describe polaritonic excitations propagating within a semiconductor waveguide. We consider the following Hamiltonian model $(\hbar=1)$
\begin{equation}\label{eq:hxc}
\begin{split}
\mathcal{\hat{H}}_{xc}=\sum_{k}&\left[\varepsilon_{x,\,k}\,\hat{X}_k^{\dagger}\hat{X}_k+ \varepsilon_{c,\,k}\hat{A}_k^{\dagger}\hat{A}_k\right.\\
&\left. +\Omega_{R}(\hat{A}_k^{\dagger}\hat{X}_k+\hat{X}_k^{\dagger}\hat{A}_k)+ V_{xx}\hat{X}_k^{\dagger}\hat{X}_k^{\dagger}\hat{X}_k\,\hat{X}_k,\right] \, ,
\end{split}
\end{equation}
which describes light-matter interactions within the propagation channel. The terms in the first line of $\mathcal{\hat{H}}_{xc}$ describe free propagating excitons and free photons, respectively. The terms in the second line account for exciton-photon interaction and exciton-exciton scattering, respectively. In particular, $\hat{X}_k^{(\dagger)}$ and $\hat{A}_k^{(\dagger)}$ denote the annihilation (creation) operators for a single exciton and a photon with wave vector $k$ into the waveguide, respectively. For such two particles, we assume the following dispersion relations
\begin{equation}
\varepsilon_{\sigma,\,k}=  \omega^{(0)}_{\sigma}+ \frac{k^2}{2\,m_{\sigma}}, \quad \sigma=x,\,c
\end{equation} 
where the first term denotes the zero-momentum energy of each excitation, and the second describing the free propagation of a particle with effective mass given by $m_{\sigma}$. Exciton-photon coupling is controlled by the Rabi frequency, $\Omega_{R}$. The parameter $V_{xx}$, which is related to the parameter $U_k$ mentioned in the main text (as detailed in the following), defines the exciton-exciton interaction energy, and it is at the origin of the polariton nonlinearity, thus playing a central role in our discussion.\\ 
In this context, exciton-polaritons emerge as the normal modes of the quadratic part of the Hamiltonian \ref{eq:hxc} annihilated by the following  operators
\begin{equation}
\hat{a}_{+,\,k}= u_k\, \hat{X}_k + v_k\, \hat{A}_k,\quad \hat{a}_{-,\,k}= -v_k\, \hat{X}_k + u_k\, \hat{A}_k,
\end{equation} 
and have the following dispersion relations 
\begin{equation}\label{eq:branches}
\omega_{\pm,\,k}= \frac{\varepsilon_{x,\,k}+\varepsilon_{c,\,k}}{2}\pm \sqrt{\left(\frac{\varepsilon_{x,\,k}-\varepsilon_{c,\,k}}{2}\right)^{2}+\Omega^{2}_{R}},
\end{equation}
with 
\begin{equation}
u_k=\frac{1}{\sqrt{1+\left(\frac{\varepsilon_{x,\,k}-\omega_{+,\,k}}{\Omega_{R}}\right)^2}},
\end{equation}
and
\begin{equation}
v_k=\frac{1}{\sqrt{1+\left(\frac{\Omega_{R}}{\varepsilon_{c,k}-\omega_{-,\,k}}\right)^2}} \, .
\end{equation}
In particular, by means of such hybrid fields, the Hamiltonian reduces to 
\begin{equation}\label{eq:hp}
\mathcal{\hat{H}}_{xc}=\sum_{k}\left[\sum_{\sigma=+,\,-}\mathcal{\hat{H}}_{\sigma,\,k}+ \hat{V}^{+,\,-}_{int,\,k}\right] \, ,
\end{equation}
with 
\begin{equation}
    \mathcal{\hat{H}}_{\sigma,\,k}=\omega_{\sigma} \hat{a}_{\sigma,\,k}^{\dagger} \hat{a}_{\sigma,\,k} +U_{\sigma,\,k}\,\hat{a}_{\sigma,\,k}^{\dagger}\hat{a}_{\sigma,\,k}^{\dagger}\hat{a}_{\sigma,\,k} \hat{a}_{\sigma,\,k} \, ,
\end{equation}
which represents the Hamiltonian operator in the branch $\sigma$, and with
\begin{equation}\label{eq:Uk}
\begin{split}
&U_{-,\,k}=v^4_k\,V_{xx}=\frac{V_{xx}}{\left[1+\left(\frac{\Omega_{R}}{\varepsilon_{x,\,k}-\omega_{+,\,k}}\right)^2\right]^2},\\
&U_{+,\,k}=u^4_k\,V_{xx}=\frac{V_{xx}}{\left[1+\left(\frac{\varepsilon_{x,\,k}-\omega_{+,\,k}}{\Omega_{R}}\right)^2\right]^2} \, ,
\end{split}
\end{equation}
denoting the interaction strength between polaritons belonging to the same branch. The last term in Eq.~\ref{eq:hp} controls interactions between polaritons belonging to different branches. \\
In general, the exciton-photon dynamics is fully described by considering both upper (+) and lower (-) polariton branches. However, if system excitations are created by shining a radiation source characterized by a spectral linewidth smaller than the exciton-photon separation, i.e., $\Delta \ll \vert\omega_c -\omega_x\vert$, the probability of populating the upper branch can be safely neglected. Therefore, in such a regime we describe the polariton field evolution by means of the following model (dropping the $``-"$ subscript)
\begin{equation}\label{eq:lower_branch_Ham}
\mathcal{\hat{H}}_L=\sum_{k}\left[\omega_{k} \hat{a}_{k}^{\dagger} \hat{a}_{k} +U_{k}\,\hat{a}_{k}^{\dagger}\hat{a}_{k}^{\dagger}\hat{a}_{k} \hat{a}_{k}\right] \, ,
\end{equation}
which represents the most common Hamiltonian model used in the literature to describe a polariton waveguide. In conclusion, if one considers two identical polariton systems described by the Hamiltonian \ref{eq:lower_branch_Ham}, and coupled to each other through an evanescent coupling (at rate $J$, as defined in the main text), it is straightforward to derive the Hamiltonian in Eq.~\ref{eq:model_Hamiltonian}.

\begin{figure}[t]
    \centering
    \includegraphics[scale=0.55]{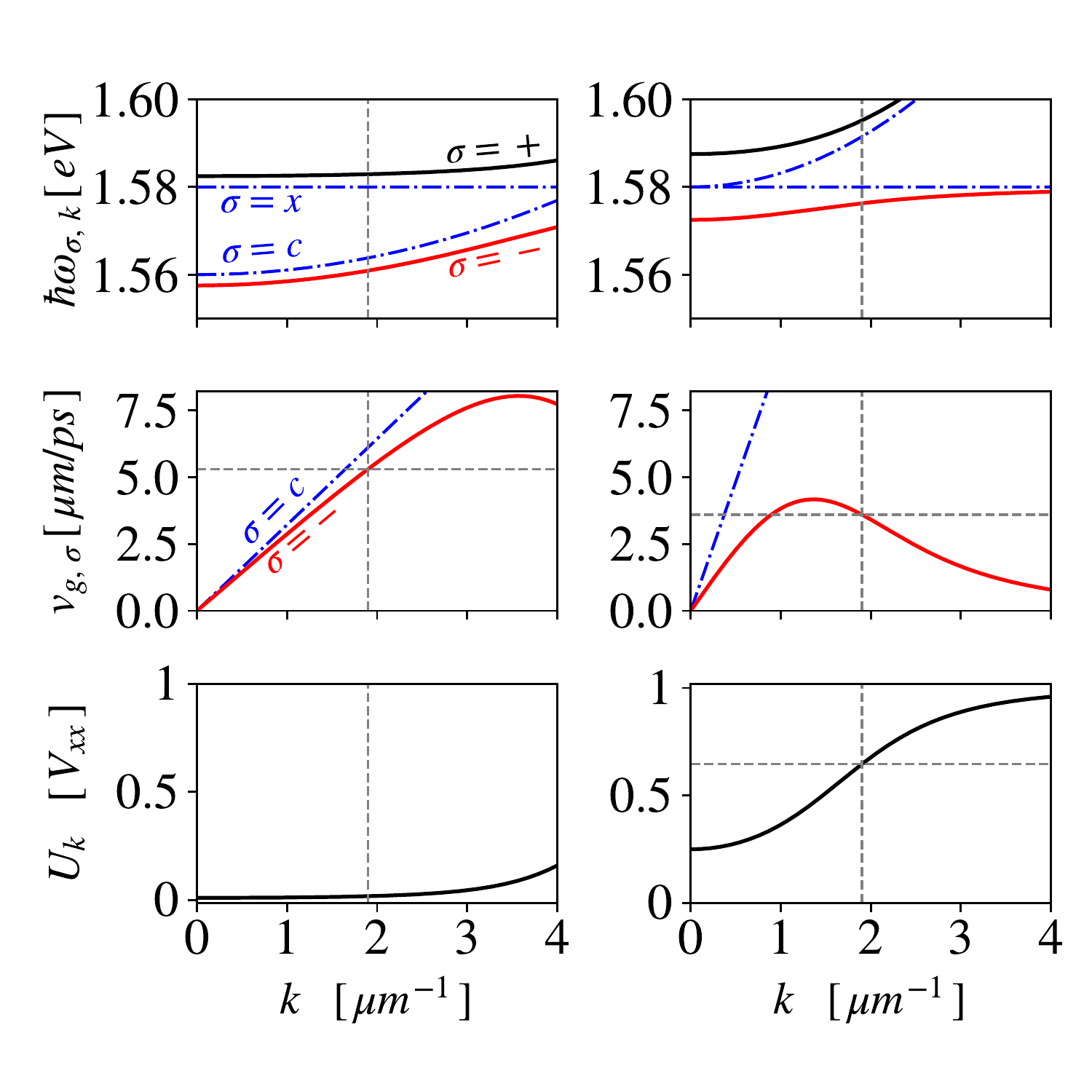}
    \caption{From the top row to the bottom one: polaritonic dispersions, group-velocity ($v_g$) and lower-branch non-linearity ($U_k$) as a function of the wavevector, $k$. Panels on the left and on the right describe the behavior obtained by means of the paramenters specified in Refs.~\cite{Beierlein2021} and \cite{Gerace2012}, respectively. For the sake of clarity and to allow for a direct comparison, we used the same value of the zero-momentum exciton energy in all the panels i.e. $\hbar\omega^{(0)}_{x}=1.58\,eV$. The horizontal (curved) dot-dashed lines in the top panels represent the exciton (photon) free dispersion relation. The vertical lines in each panel are located at $k=1.9\,\mu m ^{-1}$, which corresponds to the excitation conditions specified in Ref.~\cite{Beierlein2021}.}
    \label{fig:comparison_parameters} 
\end{figure}

\subsection{Group velocity and non-linearities}\label{app:numerelli}

 Polaritons are a mixture of photonic and excitonic degrees of freedom in pure semiconducting crystals or nanostructures, and their properties as composite particles depend on how much each of such two ingredients influences the polariton field. Here, in order to exemplify how slight changes of the system parameters can dramatically influence the global dynamics, in Fig. \ref{fig:comparison_parameters}, we report the behavior of the two polariton branches, of the group velocity $v_g=\partial\, \omega_k/\partial\,k$ and of the non-linear interaction $U_k$ as function of the propagation wavevector $k$, for the different values of the parameters used in Refs. \cite{Beierlein2021} and \cite{Gerace2012}. 
 
 \begin{table}[b]
    \centering
    \begin{tabular}{|l|c|c|}\hline
         & Ref \cite{Beierlein2021} & Ref \cite{Gerace2012}  \\
         \hline
     $\hbar\omega^{(0)}_{c}$ & 1560\, meV & 1580\, meV\\
     $\hbar\Omega_{R}$ & 7.45\,meV & 7.5\,meV \\
     $m_{c}$  & $3.6\cdot 10^{-5}\,m_0$ & $1.2\cdot 10^{-5}\,m_0$\\
     $m_{x}$  & $10^5\,m_c$ & $10^5\,m_c$ \\
    \hline
    \end{tabular}
    \caption{Parameters provided in Ref. \cite{Beierlein2021} and Ref. \cite{Gerace2012}. The parameter $m_0= 9.11\cdot 10^{-31}$ kg is the electron mass in SI units. For our purposes we set $\hbar\omega^{(0)}_{x}= 1580\, meV$ as in Ref. \cite{Gerace2012} }
    \label{tab:parameters}
\end{table}
 
 For the sake of clarity, parameters are also listed in Tab \ref{tab:parameters}. In particular, let us focus on the value $k=1.9 \mu m^{-1}$ (vertical dashed line in each panel) corresponding to the wavevector of polaritons characterized experimentally in Ref \cite{Beierlein2021}. In correspondence of such point, the scenario is qualitatively different in the two cases. As suggested by results in the left panels, in the vicinity of such $k$ polaritons of Ref \cite{Beierlein2021} essentially behave like photons: they do have a group velocity almost equivalent to that of photons in the same material (second row), and most importantly they are free propagating particles, that is $U_k\approx 0$ (third row). Data shown in the right panels describe instead particles strongly influenced by their excitonic component. Indeed, in such case for $k$ in the vicinity of $1.9$, polaritons are significantly slower than photons in the same medium (second row) and in addition they are definitely characterized by a significant non-linearity $U_k > 0.5\,V_{xx}$ (third row) .

\begin{figure}[t]
    \centering
    \includegraphics[scale=0.5]{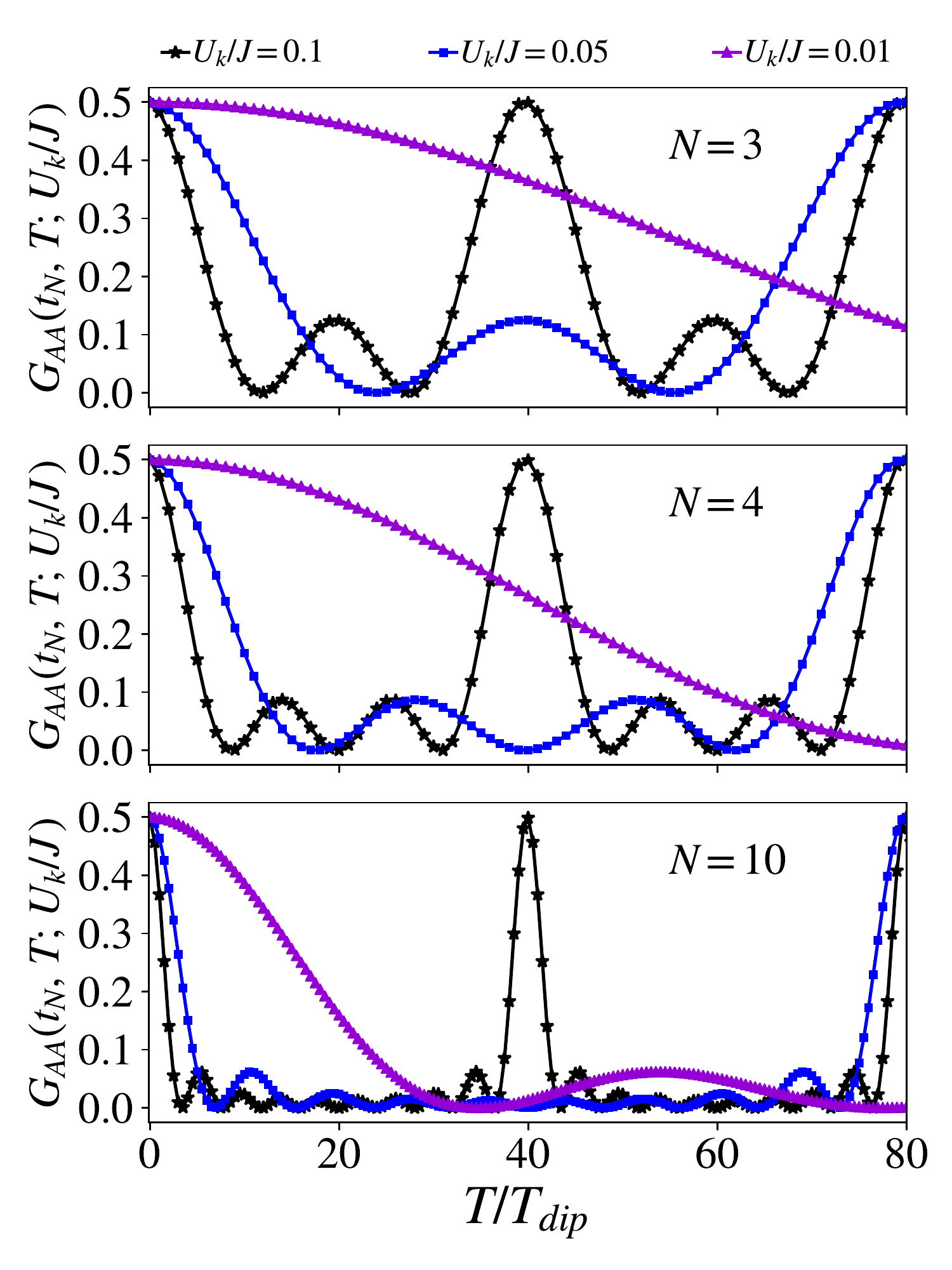}
    \caption{Behavior of $G_{AA}(t_N,\,T;\,U_k/J)$ as a function of $T/T_{dip}$, for the different values of $U_k/J$ specified on top of the figure. The parameter $t_N=T_{dip}/N$ denotes the time spent by polaritons in each of the $N$ interaction regions. From top to bottom we have $N=3,\,4,\,10$.}
    \label{fig:general_correlations}
\end{figure}

\subsection{$N$ interaction regions}\label{app:Nregions}

The analysis of the lossless QPIC is hereby generalized by considering a device having $N$ identical hopping regions separated by $N-1$ free propagation regions. By using the same notation of the main text, the action of such a device on a symmetric input is given by the following unitary operator
\begin{equation}
    U(t,\,T)= U_{IR}(t)\left(U_{FP}(T)U_{IR}(t)\right)^{N-1} \, .
\end{equation}
As a consequence, we obtain an explicit expression for the evolution operator $ U(t,\,T)$ by following the steps discussed below. First, we compute the operator $V(t,\,T)=U_{FP}(T)U_{IR}(t)$ and its $(N-1)$-th power. In particular, instead of using the representation provided in Eqs.~\ref{eq:beam_splitter_operator} and \ref{eq:free_propagation_operator}, here it is more convenient to use the exponential representation, that is
\begin{equation}
    U_{IR}(t)=e^{-i\phi t}e^{i\theta_{IR}\hat{n}_{IR}\cdot \vec{\sigma}},\quad U_{FP}(T)=e^{-i\phi T}e^{i\theta_{FP}\hat{n}_{FP}\cdot \vec{\sigma}},
\end{equation}
with $``\cdot"$ denoting the scalar product between vectors, $\phi=2\omega_k +U_k$, $\theta_{IR}=\sqrt{U_k^2+4J^2}t$, $\theta_{FP}=U_k T$, and $\vec{\sigma}$ denoting the following vector of Pauli operators
\begin{equation}
    \vec{\sigma}=\left(\sigma_{x},\,\sigma_y,\,\sigma_z\right) \, ,
\end{equation}
where $\hat{n}_{IR}$ and $\hat{n}_{FP}$ represent the following unit-length vectors 
\begin{equation}
    \hat{n}_{IR}=\left(\frac{-2J}{\sqrt{U_k^2+4J^2}},\,0,\,\frac{U_k}{\sqrt{U_k^2+4J^2}}\right),
\end{equation}
and
\begin{equation}
    \hat{n}_{FP}=\left(0,\,0,\,1\right) \, .
\end{equation}
After some algebra, we get that the $M$-th power of the product of such exponential operators reads
\begin{equation}
    V(t,\,T)^M=e^{-iM\phi (T+t)}e^{iM\theta_{v}\hat{n}_{v}\cdot \vec{\sigma}},
\end{equation}
with
\begin{equation}
    \cos(\theta_{v})=\cos(\theta_{FP})\cos(\theta_{IR})-\hat{n}_{FP}\cdot\hat{n}_{IR} \sin(\theta_{FP})\sin(\theta_{IR})
\end{equation}
and 
\begin{equation}
\begin{split}
    \hat{n}_{v} =\frac{1}{\sin(\theta_{v})} &\left[ \hat{n}_{FP}\sin(\theta_{FP})\cos(\theta_{IR}) \right.\\
    &+ \hat{n}_{IR}\cos(\theta_{FP})\sin(\theta_{IR})\\
    &-\left. \hat{n}_{FP}\times \hat{n}_{IR} \sin(\theta_{FP})\sin(\theta_{IR})\right] \, .
\end{split}
\end{equation}
As a consequence, the action of such generic device is fully encoded in the following unitary operator
\begin{equation}
    U(t,\,T)=e^{-i \phi [N( t + T) -T]} e^{i\theta_{IR}\hat{n}_{IR}\cdot \vec{\sigma}}e^{i(N-1)\theta_{v}\hat{n}_{v}\cdot \vec{\sigma}} \, .
\end{equation}

In particular, the behavior of the correlation function $G_{AA}(t=T_{dip}/N,\,T;\,U_k/J)$ as a function of $T/T_{dip}$ is shown Fig. ~\ref{fig:general_correlations}. The different curves in each panel correspond to a different value of $U_k/J$ (specified on top). The different panels correspond to a different number of interaction regions, instead: from the top panel to the bottom, we show data for $N=3,\,4,\,10$, respectively. 

Evidently, the particular values of $T/T_{dip}$ at which the probability of observing bunched photons outcoming from the device is maximized, i.e. $G_{AA}=0.5$, does not depend on $N$. Nevertheless, $N-2$ local peaks appear between the principal ones on increasing $N$, as it is clear from comparing the region $T/T_{dip}\in [0,\,40]$ in the different panels. 

\subsection{Propagation in a lossy QPIC}\label{app:losses}
In this section we account for losses in the polariton field by means of Lindblad formalism \cite{petruccione_book,Carusotto-Ciuti2013}, that is by considering the dynamics encoded into the following master equation ($\hbar=1$)
\begin{equation}\label{eq:lindblad}
\frac{d}{dt}\hat{\rho} (t) =\mathcal{L}[\rho]\equiv -i \left[\mathcal{\hat{H}},\,\hat{\rho}(t)\right]+\sum_{\sigma=a,\,b}\mathbb{D}[\hat{\sigma}]\hat{\rho} (t),
\end{equation}
with $\hat{\rho} (t)$ denoting the system density operator at time equal to $t$, and with 
\begin{equation}
    \mathbb{D}[\hat{\sigma}]\hat{\rho} (t)=\sum_{k}\Gamma\left[\hat{\sigma}_k\rho\,\hat{\sigma}_k^{\dagger}-\frac{1}{2}\left\{\hat{\sigma}_k^{\dagger}\hat{\sigma}_k,\,\hat{\rho}(t)\right\}\right],
\end{equation}
being the dissipator super operator, which accounts for losses of single polariton at rate $\Gamma$.

In the presence of such effects, the Atom-Polariton equivalence does not hold true anymore. Indeed, since losses connects sectors having a different number of polaritons, it is not possible to map the dynamics into that of a two-level Atom. Therefore, in the present scenario, in order to determine all the quantities of interest we have to deal with a $ 6 \times 6$ system of differential equations. In particular, the numerical results shown in the following sections have been obtained by performing an integration of the Lindblad equation by using a standard explicit 4-th order Runge-Kutta method with fixed time step. The time step has been chosen in such a way to ensure the numerical convergence of the expectation values. In addition, notice that the presence of losses does not alter the geometrical structure of the QPIC. This means that the propagation through a generic device like the one in Fig \ref{fig:generic_device} can be determined by concatenating the dissipative counterparts of the building blocks identified in Sec. \ref{sec:theory}.\\

In analogy with the discussion reported in the main text (see Sec. \ref{sec:theory}), here we characterize the dynamics obtained by evolving initial configurations corresponding to single polariton pure states, that is
\begin{equation}\label{eq:rho_one_A}
    \rho(t=0)= \vert 1^A_k,\,0^B_k\rangle \langle 1^A_k,\,0^B_k\vert,
\end{equation}
or 
\begin{equation}\label{eq:rho_one_B}
  \, \rho(t=0)= \vert 0^A_k,\,1^B_k\rangle \langle 0^A_k,\,1^B_k\vert,
\end{equation}
and a density operator corresponding to a couple of polaritons having the same group velocity, that is 
\begin{equation}\label{eq:initial_state_rho}
    \rho(t=0)= \vert 1^A_k,\,1^B_k\rangle \langle 1^A_k,\,1^B_k\vert. 
\end{equation}

\subsection{Decay of populations}
Let us consider first the two states in Eqs. \ref{eq:rho_one_A} and \ref{eq:rho_one_B}.  In particular, let us consider the propagation through a single component having $J(x)=J$, and let us focus on the time behavior of the populations into the two waveguides, that is $\langle \hat{n}^A_k(t)\rangle$ and $\langle \hat{n}^B_k(t)\rangle$, with $\langle \hat{O}(t)\rangle\equiv\mbox{Tr}[\hat{O}\,\rho(t)]$. By using the Heisenberg representation \cite{petruccione_book}, one finds that such quantities can be found by determining the solution of the following differential system 
\begin{equation}
    \frac{d}{dt}\left(\begin{matrix}
    \langle \hat{n}^A_k(t)\rangle\\
    \langle \hat{n}^B_k(t)\rangle\\
    \langle \hat{b}^{\dagger}_k\hat{a}_k(t)\rangle\\
    \langle \hat{a}^{\dagger}_k\hat{b}_k(t)\rangle\\
    \end{matrix}\right)=
    \left(\begin{matrix}
    -\Gamma & 0 & iJ & -iJ \\
    0 & -\Gamma &-iJ & iJ \\
    iJ & -iJ & -\Gamma & 0 \\
    -iJ & iJ & 0 & -\Gamma\\
    \end{matrix}
    \right)\left(\begin{matrix}
    \langle \hat{n}^A_k(t)\rangle\\
    \langle \hat{n}^B_k(t)\rangle\\
    \langle \hat{b}^{\dagger}_k\hat{a}_k(t)\rangle\\
    \langle \hat{a}^{\dagger}_k\hat{b}_k(t)\rangle\\
    \end{matrix}\right),
\end{equation}
with the following initial condition in the case of Eq. \ref{eq:rho_one_A}
\begin{equation}\label{eq:initial_one_A}
    \langle \hat{n}^A_k(0)\rangle=1,\,    \langle \hat{n}^B_k(0)\rangle=\langle \hat{b}^{\dagger}_k\hat{a}_k(0)\rangle=\langle \hat{a}^{\dagger}_k\hat{b}_k(0)\rangle=0,\,
\end{equation}
and with initial condition given by 
\begin{equation}\label{eq:initial_one_B}
    \langle \hat{n}^B_k(0)\rangle=1,\,    \langle \hat{n}^A_k(0)\rangle=\langle \hat{b}^{\dagger}_k\hat{a}_k(0)\rangle=\langle \hat{a}^{\dagger}_k\hat{b}_k(0)\rangle=0,\,
\end{equation}
if $\rho(t=0)$ is given by the state in Eq. \ref{eq:rho_one_B}.\\ After a little algebra, one finds that the solution of such differential system when considering the initial condition provided in Eq. \ref{eq:initial_one_A} gives the following behavior for the populations
\begin{equation}\label{eq:solution_initial_A}
    \langle \hat{n}^A_k(t)\rangle= e^{-\Gamma\,t}\cos^{2}(Jt),\quad \langle \hat{n}^B_k(t)\rangle= e^{-\Gamma\,t}\sin^{2}(Jt).
\end{equation}
A similar computation shows that the initial condition in Eq. \ref{eq:initial_one_B} give instead the following behavior
\begin{equation}\label{eq:solution_initial_B}
    \langle \hat{n}^A_k(t)\rangle= e^{-\Gamma\,t}\sin^{2}(Jt),\quad \langle \hat{n}^B_k(t)\rangle= e^{-\Gamma\,t}\cos^{2}(Jt),
\end{equation}
where the roles of the two waveguides have been exchanged with respect to the behaviors reported in Eq. \ref{eq:solution_initial_A}, due to the different initial conditions. Interestingly, a comparison of such results with those shown in the main text for the same observables (see Eq. \ref{eq:BS_action} and \ref{eq:BS_times}) reveals that the 50:50 beamsplitting condition is not affected by losses.\\

Let us consider now as initial condition the state in Eq. \ref{eq:initial_state_rho}. In this case, in order to determine the time behavior of populations, we take advatage of the system symmetries. The Lindblad superoperator is weakly symmetric \cite{Baumgartner2008,Buca2012,Albert2014,Nigro2020} under the exchange of the two operators $\hat{a}_k$ and $\hat{b}_k$, i.e. it satisfies 
\begin{equation}\label{eq:weak_symmetry}
    U_{\hat{a}\leftrightarrow \hat{b}} \mathcal{L}[\rho(t)]U^{\dagger}_{\hat{a}\leftrightarrow \hat{b}}=\mathcal{L}   \left[U_{\hat{a}\leftrightarrow \hat{b}} \rho(t)U^{\dagger}_{\hat{a}\leftrightarrow \hat{b}} \right], 
\end{equation}
with $U_{\hat{a}\leftrightarrow \hat{b}}$ denoting the unitary operator that exchanges polaritons in different waveguides. As a consequence, by noting that the following vectors
\begin{equation}
\begin{split}
\left\{\vert v^{(+)}_j\rangle\right\}\equiv&  \left\{\vert \Omega\rangle,\,\frac{\vert 1^{A}_k,0\rangle+\vert 0,\, 1^{B}_k\rangle}{\sqrt{2}},\right.\\
&\quad\left.\vert 1^{A}_{k},\,1^{B}_{k}\rangle,\,\frac{\vert 2^{A}_{k},\,0\rangle +\vert 0,\,2^{B}_{k}\rangle}{\sqrt{2}}\right\},\\
\end{split}
\end{equation}
and
\begin{equation}
\left\{\vert v^{(-)}_j\rangle\right\}\equiv  \left\{\frac{\vert 1^{A}_k,0\rangle-\vert 0,\, 1^{B}_k\rangle}{\sqrt{2}},\,\frac{\vert 2^{A}_{k},\,0\rangle - \vert 0,\,2^{B}_{k}\rangle}{\sqrt{2}}\right\},
\end{equation}
are eigenstates of $U_{\hat{a}\leftrightarrow \hat{b}}$ with eigenvalues $\lambda=+1$ and $\lambda=-1$ respectively, that is $U_{\hat{a}\leftrightarrow \hat{b}}\vert v^{(\pm)}_j\rangle=\pm \vert v^{(\pm)}_j\rangle$, the density operator can be decomposed in this basis as the superposition of a symmetric term and a skew-symmetric one, that is 
\begin{equation}
    \rho(t)= \rho^{(+)}(t)+ \rho^{(-)}(t),
\end{equation}
with $U_{\hat{a}\leftrightarrow \hat{b}} \rho^{(\pm)}(t)U^{\dagger}_{\hat{a}\leftrightarrow \hat{b}}=\pm \rho^{(\pm)}(t)$. Most importantly, following from Eq. \ref{eq:weak_symmetry}, the time evolution encoded into the Lindblad master equation can be decomposed as the sum of two independent terms, a symmetric and skew-symmetric part, that is
\begin{equation}
    \frac{d\rho^{(\pm)}(t)}{dt}=\mathcal{L}[\rho^{(\pm)}(t)].
\end{equation}
In the present case, since  the initial configuration in Eq. \ref{eq:initial_state_rho} is symmetric, we do have that such property is preserved for any time $t>0$. Furthermore, since any given observable $\hat{O}$ can always be decomposed as a symmetric and a skew-symmetric part, that is 
\begin{equation}
    \hat{O}=\hat{O}^{(+)}+ \hat{O}^{(-)}, \quad U_{\hat{a}\leftrightarrow \hat{b}}\hat{O}^{(\pm)}U^{\dagger}_{\hat{a}\leftrightarrow \hat{b}}=\pm \hat{O}^{(\pm)},
\end{equation}
in the present case we do have that 
\begin{equation}\label{eq:symm_ops}
    \langle \hat{O} (t)\rangle \equiv\mbox{Tr}[\hat{O} \rho(t)]=\mbox{Tr}[\hat{O}^{(+)} \rho(t)]=\langle \hat{O}^{(+)} (t)\rangle \, . 
\end{equation}
Specifying to our parallel waveguides case, we conclude that the average number of polaritons at any time $t$ is independent of the A or B channel, as in the lossless regime, i.e.,
\begin{equation}
\begin{split}
    \langle \hat{n}^{A}_k (t)\rangle &= \mbox{Tr}\left[  \hat{n}^A_k\,\rho(t)\right]=\\ &=\mbox{Tr}\left[  \hat{n}^A_k\,U_{\hat{a}\leftrightarrow \hat{b}}\rho(t)U^{\dagger}_{\hat{a}\leftrightarrow \hat{b}}\right]=\langle \hat{n}^B_k (t)\rangle \, ,
    \end{split}
\end{equation}
which is also valid for second-order observables, such as  
\begin{equation}
G_{AA}(t;\,U_k/J,\,\Gamma/J)=G_{BB}(t;\,U_k/J,\,\Gamma/J),
\end{equation}
with $G_{AA}(t;\,U_k/J,\,\Gamma/J)=\mbox{Tr}\left[a^{\dagger}_k\hat{a}^{\dagger}_k \hat{a}_k\hat{a}_k\,\rho(t)\right]$, where the dependence on $\Gamma/J$ is implicitly hidden in the density operator $\rho(t)$.\\
Consider now the Heisenberg equation of motion for the operator $\hat{n}_A$, that is
\begin{equation}
    \frac{d\hat{n}^{A}_k}{dt}=iJ(\hat{b}^{\dagger}_k\hat{a}_k -\hat{a}^{\dagger}_k\hat{b}_k)-\Gamma\hat{n}^A_k.
\end{equation}
Due to the weak symmetry of the Lindblad equation, to the relation in Eq.~\ref{eq:symm_ops} and to the form of the initial condition, we have that $\langle \hat{b}^{\dagger}_k\hat{a}_k -\hat{a}^{\dagger}_k\hat{b}_k\rangle$ is identically zero at any $t$. Therefore, in the present case one has that 
\begin{equation}\label{eq:number_decay}
    \langle \hat{n}^A_k (t)\rangle = e^{-\Gamma  t}=  \langle \hat{n}^B_k (t)\rangle \, ,
\end{equation}
which confirms that by considering a master equation in the form provided in Eq.~\ref{eq:lindblad} leads to the behavior expected for the polariton population, that is a decaying behavior at a rate $\Gamma$.

\begin{figure}[t]
    \centering
    \includegraphics[scale=0.5]{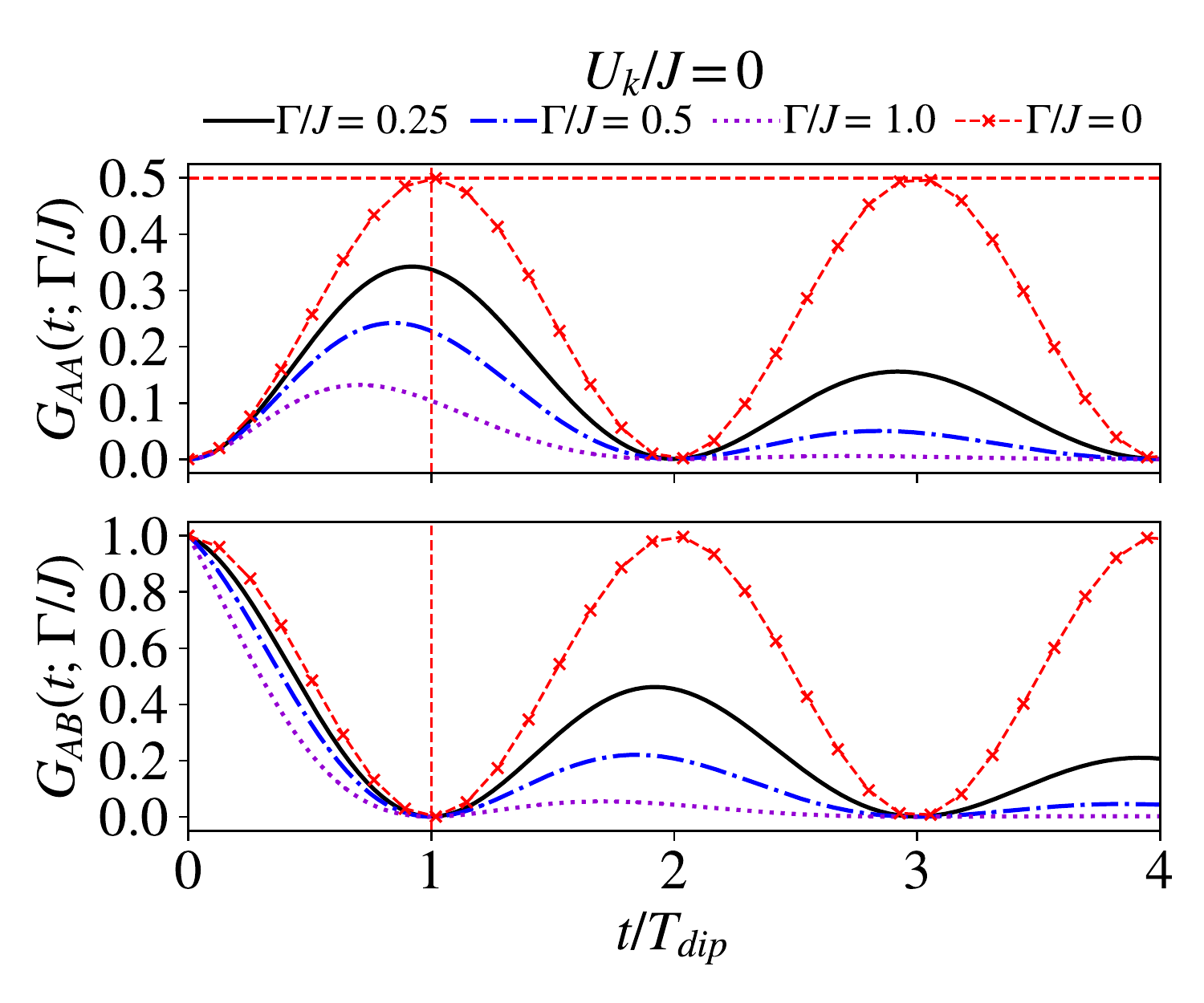}
    \caption{Dependence of $G_{AA}(t;\,\Gamma/J)\equiv G_{AA}(t;U_k/J=0,\,\Gamma/J)$ and $G_{AB}(t;\,\Gamma/J)\equiv G_{AB}(t;U_k/J=0,\,\Gamma/J)$ on $t/T_{dip}$, for $U_k/J=0$ and for different values of $\Gamma/J$. The horizontal and vertical dashed lines have been added as a guide to the eye evidencing the HOM condition along the $t/T_{dip}$ axis, and the HOM maximum of $G_{AA}(t=T_{dip};\,U_k=0,\,\Gamma/J=0)=0.5$.}
    \label{fig:decaying_correlations_HOM}
\end{figure}

\subsection{Single hopping region}

In this Section we address the issue of understanding how losses might affect correlations between photons emitted by the device depicted in Fig.~\ref{fig:HOM_setup}. Indeed, in realistic situations the polariton population within each waveguide will decay due to their finite lifetime. From a practical point of view, then, the waveguide length have to allow polaritons to reach the detection region.\\
As it has been mentioned in Sec.~\ref{sec:theory}, here we take into account any loss channel that can be theoretically described by the Lindblad master equation in \ref{eq:lindblad}. In particular, it has already been shown in the section before that, by choosing as an initial state the density operator in Eq. \ref{eq:initial_state_rho}, the master equation leads to a the time-dependent average number of polaritons having the following analytic expression
\begin{equation}\label{eq:number_decay}
    \langle \hat{n}^{A}_k (t)\rangle =  e^{-\Gamma  t} = \langle \hat{n}^{B}_k (t)\rangle \, .
\end{equation}
Notice, in particular, that the behavior in Eq.~\ref{eq:number_decay} does not depend on $U_k/J$, but only on the particular value of $\Gamma$. As in the lossless regime, also in this case we begin our analysis of correlation functions by considering the un-normalized expectation values $G_{AA}(t;\,U_k/J,\,\Gamma/J)=\mbox{Tr}\left[a^{\dagger}_k\hat{a}^{\dagger}_k \hat{a}_k\hat{a}_k\,\rho(t)\right]$ and $G_{AB}(t;\,U_k/J,\,\Gamma/J)=\mbox{Tr}\left[a^{\dagger}_k\hat{b}^{\dagger}_k \hat{b}_k\hat{a}_k\,\rho(t)\right]$ (we remind that $G_{BB}(t;\,U_k/J,\,\Gamma/J)=G_{AA}(t;\,U_k/J,\,\Gamma/J)$). However, notice that contrary to the unitary case, due to the decaying behavior of the total number of particles, these two expectation values do not correspond to the proper observables suitable for the characterization of the degree of correlation of photons emitted by a realistic lossy device. To this purpose, in Fig.~\ref{fig:decaying_correlations_HOM} we show the behavior of these two quantities for $U_k/J=0$, and for different values of $\Gamma /J$ (specified on top of the figure), as a function of $t/T_{dip}$.  

\begin{figure}[t]
    \centering
    \includegraphics[scale=0.5]{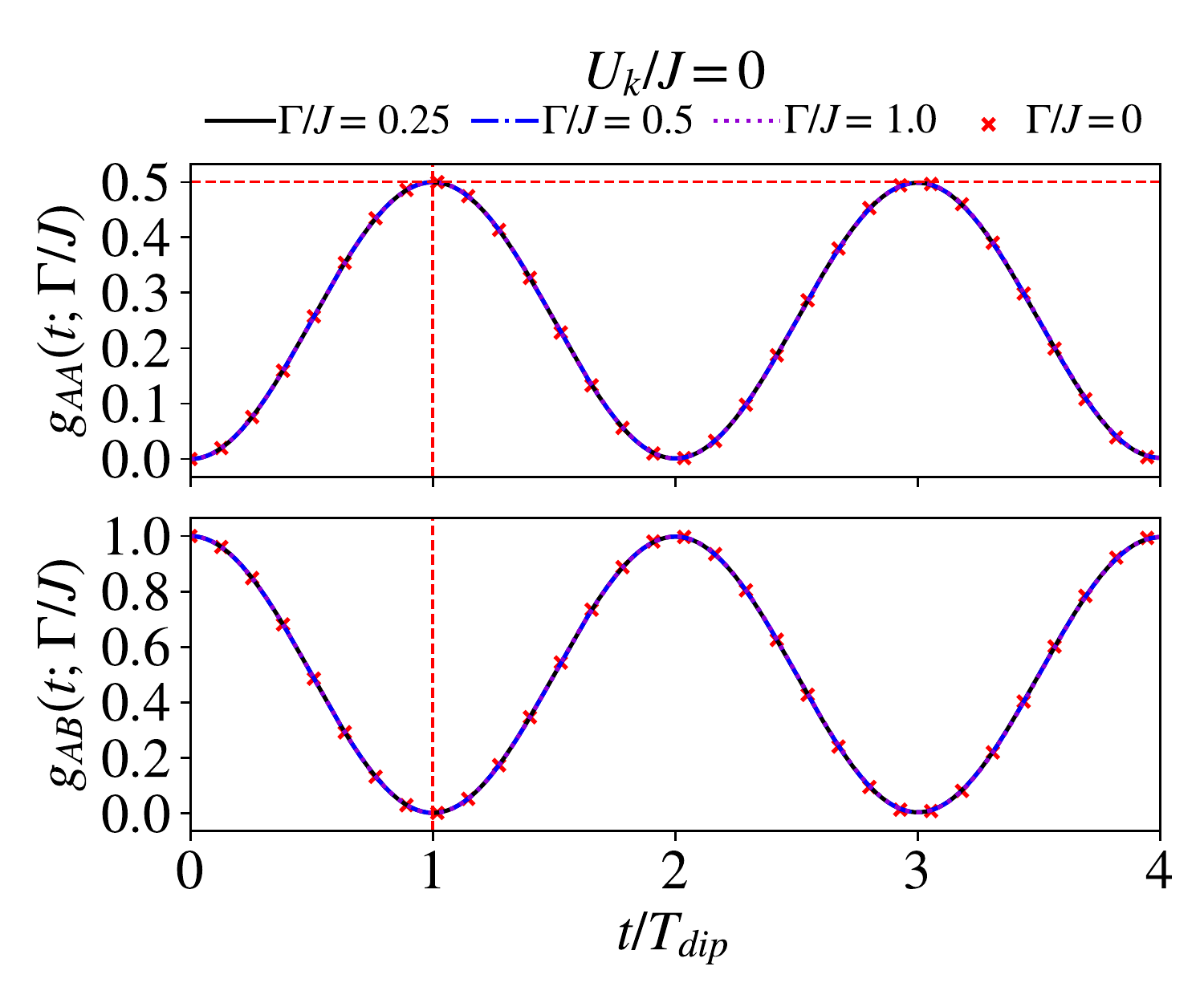}
    \caption{Dependence of normalized correlation functions, $g_{AA}(t;U_k/J,\,\Gamma/J)$ and of $g_{AB}(t;U_k/J,\,\Gamma/J)$, for $U_k/J=0$ and different values of $\Gamma/J$, as a function of $t/T_{dip}$, as calculated from the numerical solution of the quantum master equation. The different curves are not distinguishable. The points represent the analytic result obtained in the previous Section, and corresponding to the lossless case, $\Gamma/J=0$. }
    \label{fig:normalized_correlations_loss_HOM}
\end{figure}

Evidently, such amplitudes decay very fast while increasing the interaction length (i.e., $t/T_{dip}$), as soon as $\Gamma/J$ increases. However, notice that this means only that while increasing both the losses and the length of the device, it becomes more and more unlikely to have a pair of photons simultaneously reaching the detection region, either for what concerns photons coming from the same waveguide (top panel of Fig.~\ref{fig:decaying_correlations_HOM}) or photons coming from different channels (bottom panel). Indeed, in order to properly detect correlations, and most importantly to understand if losses might modify the correlation behavior discussed in the previous Sections, it is crucial to plot normalized amplitudes by considering the proper number of photodetection events occurring at each instant $t$, i.e. the quantities
\begin{equation}\label{eq:normalized_N_AA_HOM}
g_{AA}(t;U_k/J,\,\Gamma/J)=\frac{G_{AA}(t;U_k/J,\,\Gamma/J)}{\langle \hat{n}^{A}_k (t)\rangle \langle \hat{n}^{A}_k (t)\rangle} \, ,
\end{equation}
and 
\begin{equation}\label{eq:normalized_N_AB_HOM}
g_{AB}(t;U_k/J,\,\Gamma/J)=\frac{G_{AB}(t;U_k/J,\,\Gamma/J)}{\langle \hat{n}^{A}_k (t)\rangle \langle \hat{n}^{B}_k (t)\rangle}.
\end{equation}
The numerators of the expressions in Eqs.~\ref{eq:normalized_N_AA_HOM} and \ref{eq:normalized_N_AB_HOM} count double click events occurring on the same detector and simultaneously on different detectors, respectively. On the other hand, the denominators of the expressions in Eqs.~\ref{eq:normalized_N_AA_HOM} and \ref{eq:normalized_N_AB_HOM} count how many photon pairs emerge from the same waveguide channel and from different channels, respectively, at time $t$. Interestingly, such normalized quantities reveal that the correlations are not affected by losses. This is explicitly shown in Fig.~\ref{fig:normalized_correlations_loss_HOM}, where the same data reported in Fig.~\ref{fig:decaying_correlations_HOM} have been normalized according to the definitions provided into Eqs.~\ref{eq:normalized_N_AA_HOM} and \ref{eq:normalized_N_AB_HOM}. Here the different curves corresponding to different decay rates $\Gamma/J$, which are clearly distinguishable in Fig.~\ref{fig:decaying_correlations_HOM}, are now indistinguishably given on the same reference curve. In particular, such a reference curve is exactly the one previously determined in the lossless regime (plotted with points and corresponding to the condition $\Gamma/J=0$). 

\begin{figure}[t]
    \centering
    \includegraphics[scale=0.5]{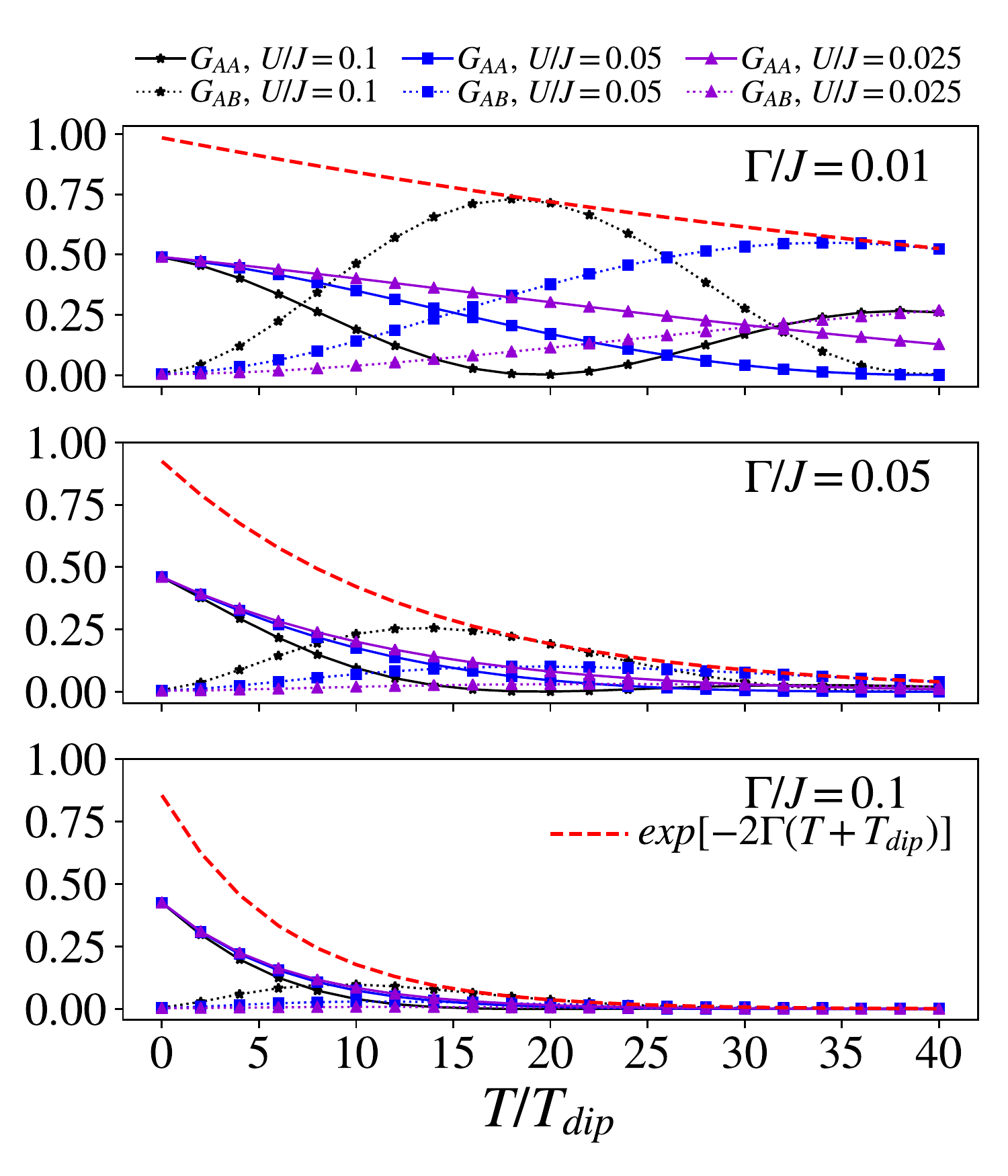}
    \caption{Dependence of $G_{AA}\equiv G_{AA}(t;\,U_k/J,\,\Gamma/J)$ (solid lines) and of $G_{AB}\equiv G_{AB}(t;\,U_k/J,\,\Gamma/J)$ (dotted lines) on $T/T_{dip}$. The different curves correspond to different values of the ratio $U_k/J$ (specified in the legend). The data plotted in the three panels describe the behavior of such correlation functions for the different values of $\Gamma/J$ specified in each panel. The red dashed line in each panel  represents the time dependence of $\langle n^{A}_{k}(T+T_{dip})\rangle^2=\exp(-2\Gamma (T+T_{dip})$.}
    \label{fig:RAMSEY_dissipative_correlations}
\end{figure}

Furthermore, the same occurs also for $U_k/J \neq 0$. In particular, by studying the numerical results for $U_k/J>0$ (not shown), we also find that the correlation functions in the presence of losses are compatible with the analytic expressions
\begin{equation}
    G_{AA}(t;\,U_k/J,\,\Gamma/J)=e^{-2\Gamma t} G_{AA}(t;\,U_k/J,\,\Gamma/J=0),
\end{equation}
and  
\begin{equation}
    G_{AB}(t;\,U_k/J,\,\Gamma/J)=e^{-2\Gamma t}G_{AB}(t;\,U_k/J,\,\Gamma/J=0) \, .
\end{equation}
In other words, if on the one hand the presence of losses inevitably put some constraints on the device structure, on the other hand the behavior of normalized correlation functions that could be measured from photon counting experiments reveal that losses do not change the conclusions previously reached in Sec. \ref{subsec:HOM_lossless} for the lossless case.

\begin{figure}[t]
    \centering
    \includegraphics[scale=0.5]{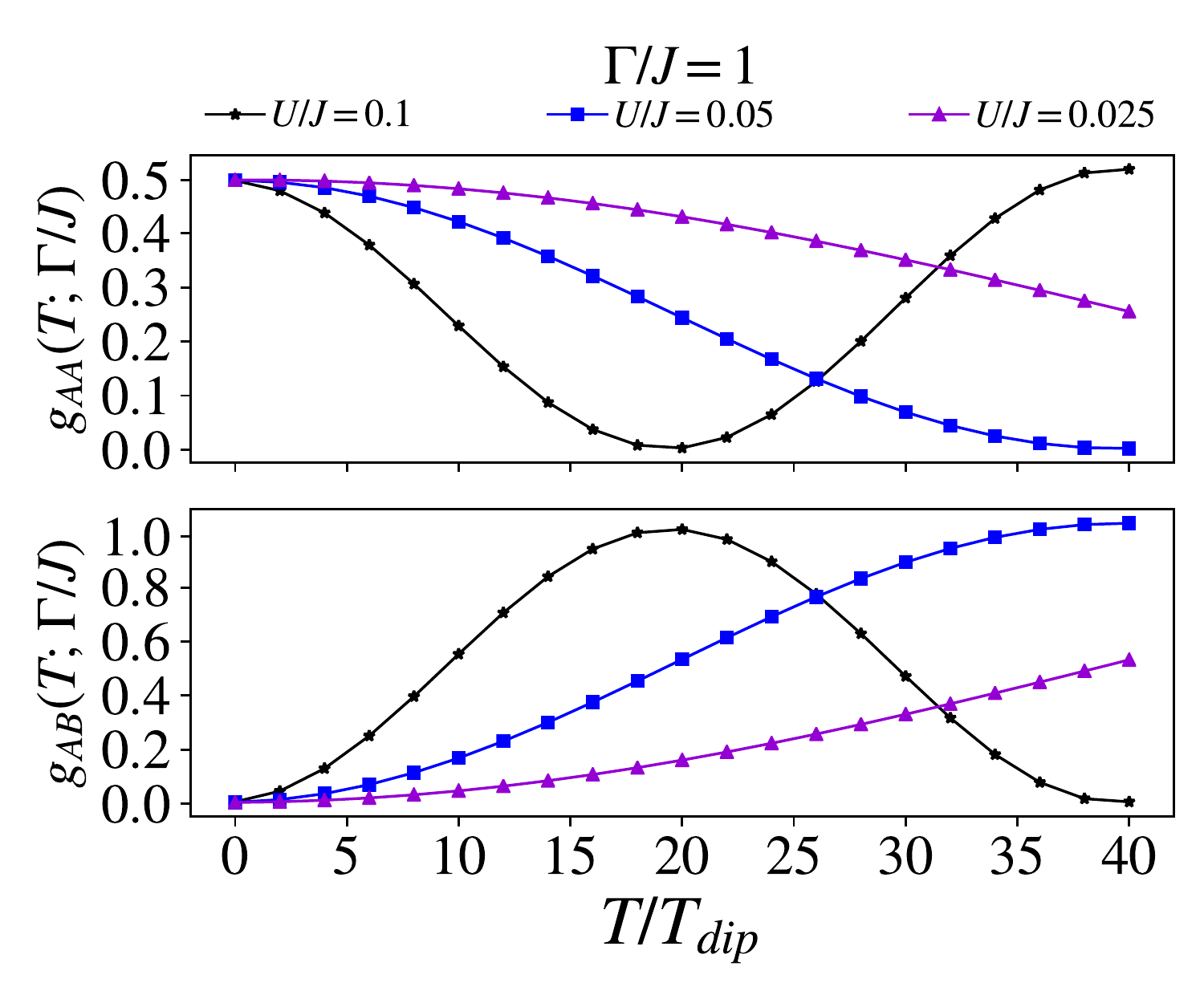}
    \caption{Behavior of $g_{AA}(t=T_{dip}/2,\,T;\,U_k/J,\,\Gamma/J)\equiv g_{AA}(T;\Gamma/J)$ and of $g_{AB}(t,\,T;\,U_k/J,\,\Gamma/J)\equiv g_{AB}(T;\Gamma/J)$ for $\Gamma/J=1$ and different values of $U_k/J$, as a function of $T/T_{dip}$.}
    \label{fig:RAMSEY_dissipative_correlations_NORMALIZED}
\end{figure}

\subsection{Two hopping regions:\\ the lossy Ramsey interferometer}

In this section, we address the issue of losses and how they affect the measured correlation functions of photons emerging from the device in Fig.~\ref{fig:Ramsey_setup}. As in the previous Section, here we assume the system to be initially prepared in a state described by the density operator defined in Eq.~\ref{eq:initial_state_rho}. First, we show the numerical solutions for $G_{AA}(t;\,U_k/J,\,\Gamma/J)$ and $G_{AB}(t;\,U_k/J,\,\Gamma/J)$ corresponding to the condition $t/T_{dip}=0.5$, for different values of $U_k/J$ and $\Gamma/J$, respectively. The results are shown in Fig.~\ref{fig:RAMSEY_dissipative_correlations}, where we see that, in analogy to the HOM setup, it is more and more unlikely to measure coincident photons emerging at the output, on increasing $\Gamma/J$. In particular, similarly to the previous case, the decaying behavior is compatible with the exponential decay of $\langle n^{A}_{k}(T)\rangle^2$ (again, we remind that $\langle n^{A}_{k}(T)\rangle=\langle n^{B}_{k}(T)\rangle$), for any given value of $\Gamma/J$. In Fig.~\ref{fig:RAMSEY_dissipative_correlations} the population decay is plotted with red dashed curves in each panel. Also in this case, the normalized correlation functions defined in Eqs.~\ref{eq:normalized_N_AA_HOM} and \ref{eq:normalized_N_AB_HOM} behave as the correlation functions in the absence of losses, i.e., as the functions defined in Eqs.~\ref{eq:G_AA_Ramsey} and \ref{eq:G_AB_Ramsey}. An example of such one-to-one correspondence is provided by the numerical results shown in Fig.~\ref{fig:RAMSEY_dissipative_correlations_NORMALIZED}, where we show the behavior of normalized correlations for $\Gamma/J=1$, i.e., in the presence of strong dissipation, and for different values of $U_k/J$.  

Interestingly, after proper normalization the data for $\Gamma/J=1$ behave exactly as those obtained in the lossless regime, as it is evident by comparing  Figs.~\ref{fig:RAMSEY_dissipative_correlations_NORMALIZED} and  \ref{fig:Ramsey_Correlations_Cut}. We found the same correspondence, that is the collapse of all the data for a given $U_k/J$ and different values of $\Gamma/J$ to the curve obtained for $\Gamma/J=0$, also by using the numerical results reported in Fig.~\ref{fig:RAMSEY_dissipative_correlations} (not shown). Therefore we infer that 
\begin{equation}
\begin{split}
    G_{AA}(t,&\,T;\,U_k/J,\,\Gamma/J)=\\
    &=e^{-2\Gamma (2t+T)} G_{AA}(t,\,T;\,U_k/J,\,\Gamma/J=0),\\
\end{split}
\end{equation}
and  
\begin{equation}
\begin{split}
    G_{AB}(t,&\,T;\,U_k/J,\,\Gamma/J)=
    \\
    &=e^{-2\Gamma (2t+T)}G_{AB}(t,\,T;\,U_k/J,\,\Gamma/J=0) \, ,
\end{split}
\end{equation}
similarly to what we already obtained for a single hopping region in the presence of losses, at the end of the previous subsection.

\bibliography{bibliography}

\end{document}